%% file: main.tex
\documentclass[11pt,letterpaper]{article}
\usepackage[margin=1in]{geometry}
\usepackage{microtype}
\usepackage[utf8]{inputenc}
\usepackage{graphicx}
\usepackage{tikz}
\usepackage{tabularx}
\usetikzlibrary{arrows.meta,decorations.pathreplacing}
\usepackage{gensymb}
\usepackage{enumitem}
\usepackage{amsmath}
\usepackage{amssymb}
\usepackage{amsthm}
\usepackage{booktabs}
\usepackage{xcolor}
\usepackage{todonotes}

\usepackage[colorlinks,allcolors=blue]{hyperref}
\usepackage{cleveref}
\usepackage{thmtools}
\usepackage{thm-restate}
\newtheorem{theorem}{Theorem}
\newtheorem{lemma}[theorem]{Lemma}
\newtheorem{meta-theorem}[theorem]{Meta-Theorem}

\newtheorem{observation}[theorem]{Observation}

\theoremstyle{remark}
\newtheorem*{remark}{Remark}

\newcommand{\mwcs}{\operatorname{mwcs}}
\newcommand{\PP}{\mathcal{P}}
\newcommand{\T}{\mathcal{T}}

\definecolor{redpt}{RGB}{215,35,35}
\definecolor{bluept}{RGB}{25,80,205}
\definecolor{tA}{RGB}{0,150,95}
\definecolor{tB}{RGB}{235,145,0}
\definecolor{tC}{RGB}{140,75,185}
\definecolor{tD}{RGB}{0,165,180}
\definecolor{tE}{RGB}{150,100,55}
\tikzset{
  rp/.style={circle,fill=redpt,inner sep=0pt,minimum size=4.4pt},
  bp/.style={circle,fill=bluept,inner sep=0pt,minimum size=4.4pt},
  rps/.style={circle,fill=redpt,inner sep=0pt,minimum size=3.2pt},
  bps/.style={circle,fill=bluept,inner sep=0pt,minimum size=3.2pt},
  rpo/.style={circle,draw=redpt,thick,fill=white,inner sep=0pt,minimum size=5pt},
  bpo/.style={circle,draw=bluept,thick,fill=white,inner sep=0pt,minimum size=5pt},
  lab/.style={font=\scriptsize},
}

\begin{document}
\title{Near-Linear Time Maximum Weight Clique for Objects Inscribed in a Circle}
\author{Jaehyun Koo \\ \small Massachusetts Institute of Technology \\ \small koosaga@mit.edu \and
Dario Moschetti \\ \small University of Milan \\ \small dario.moschetti@studenti.unimi.it}
\date{}

\maketitle

\begin{abstract}
Circle trapezoid graphs are the intersection graphs of regions of a disk between two non-crossing chords, and they contain circle graphs, circular-arc graphs and trapezoid graphs. We give an $O(n\log^2 n)$-time algorithm for maximum weight clique in circle trapezoid graphs, improving on the $O(n^2\log n)$ bound known since 1997, and an $O(N\log^2 N)$-time algorithm for polygon-circle graphs, where $N$ is the total number of vertices of the polygons. This gives the first worst-case subquadratic algorithms for maximum weight clique in circle graphs and in circular-arc graphs, answering questions left open in previous work. All these classes are handled by one reduction to a dominance problem on red and blue points in the plane. Unlike the previous near-linear time algorithms for circle graphs and circular-arc graphs, ours do not use Monge matrices: they rely on elementary dynamic programming, which also gives a simpler alternative in the unweighted case. From the same approach we also obtain a data structure for interval queries on red and blue points, a more general problem that contains weighted range longest increasing subsequence (LIS) queries, matching the best known bounds for the latter.
\end{abstract}

\section{Introduction}

Finding a maximum clique in a graph is one of Karp's 21 NP-complete problems \cite{Karp72}, and a classical line of work asks for which restricted graph classes the problem becomes tractable. Geometric intersection graphs are a natural candidate: every vertex represents a geometric object, two vertices are adjacent when their objects intersect, and the input is usually the list of the $n$ objects, so that the graph is only implicit. The hope is that the geometry of the objects can be exploited to solve the problem efficiently. This does not always work, as for some objects the problem remains NP-hard \cite{CCL13, AW05}, but for many others it leads to efficient algorithms \cite{GLL82, IA83, CCJ90}.

In this paper we study the problem for a family of graph classes whose objects live in a disk and touch its boundary circle: arcs of the circle (\emph{circular-arc graphs}), chords of the circle (\emph{circle graphs}), regions between two non-crossing chords (\emph{circle trapezoid graphs} \cite{FMW97}), and convex polygons inscribed in the circle (\emph{polygon-circle graphs} \cite{Koe92,KK97}). This condition alone does not make the problem tractable: maximum clique is NP-hard already for segments in the disk with one endpoint on the circle (\emph{outer segment graphs}) \cite{CCL13,KMMN22}.

\subsection{Related work}

Circle graphs and circular-arc graphs come with a long history of clique algorithms, and in both, the weighted version, where every object has a weight and we look for a clique of maximum total weight, has resisted the techniques that work for the unweighted version.

\paragraph{Circle graphs.}
The first polynomial time algorithm to find a maximum clique in a circle graph was proposed by Gavril \cite{Gav73}. A line of improvements \cite{RU81, Hsu85, MNKF90, AAH93} brought the time down to $O(n\log n+\min\{m,\,n\omega\})$ for the maximum clique \cite{MNKF90} and $O(n\log n+\min\{n^2,\,m\log\log n\})$ for the maximum weight clique \cite{AAH93}, where $m$ is the number of edges of the implicit graph and $\omega$ the clique number. All these bounds are quadratic in the worst case. Tiskin was the first to break the quadratic barrier for the unweighted version, through an elegant connection to a string problem that is in turn solved by multiplying \emph{unit-Monge matrices}: his algorithm runs in $O(n^{1.5})$ time \cite{Tis06}, later improved to $O(n\log^2 n)$ \cite{Tis15}. The 2006 paper also asked whether the method extends to the weighted problem \cite[Section~6]{Tis06}.

\paragraph{Circular-arc graphs.}
For the unweighted version of the problem, after a line of improvements from an initial $O(nm)$ running time \cite{Hsu85,AH87,SH89}, Bhattacharya and Kaller gave an algorithm running in $O(n\log n+m)$ time \cite{BK97}, and asked whether $O(n\log n)$ was possible. The question was repeated by Tiskin \cite{Tis06} and recently answered, up to a logarithmic factor, by Moschetti, Polak and Schmidt \cite{MPS27}, who solved the problem in $O(n\log^2 n)$ time. For the weighted version, Shih and Hsu gave an algorithm running in $O(n\log n+m\log\log n)$ time \cite{SH89}, later matched in \cite{BK97} and refined in \cite{MPS27} to $O(dn\log\log n+n\log n)$, where $d\le 2m/n$ is the minimum load, the smallest number of arcs containing a point of the circle. That paper also asked whether maximum weight clique in circular-arc graphs can be solved in subquadratic time.

\paragraph{Monge matrices and weights.}
The two subquadratic algorithms above, for circle graphs \cite{Tis06,Tis15} and for circular-arc graphs \cite{MPS27}, rely on the same tool: in two different ways, they encode the problems in Monge matrices with a sparse structure, and multiply these matrices fast. The tool comes from string algorithms. Tiskin observed that the Monge matrices arising from unweighted longest common subsequence (LCS), longest increasing subsequence (LIS) and edit distance, and from cliques in circle graphs, are \emph{unit-Monge} and admit linear-size representations \cite{Tis06,Tis08,Tis08a,Tis15}. Russo later generalized this notion to \emph{core-sparse} Monge matrices, in which the \emph{core}, the set of nonzero entries of the density matrix, is small \cite{Rus12}, allowing weights and relaxing some structural requirements. Fast multiplication of unit-Monge and core-sparse Monge matrices \cite{Tis15,GGK25} has since been used for approximate pattern matching \cite{CKW22,CKW25}, and for dynamic \cite{CKM20,KS21,GK25a}, parallel \cite{MBT21,CHS23} and distributed \cite{KT10a,Koo24} algorithms for problems admitting such representations.

The main drawback of this approach is that it is hard to extend to weights. With weights, the matrices remain Monge, but their cores are no longer guaranteed to be small, and no subquadratic bound on their size is known. For circular-arc graphs, such a bound would already give, through the reduction of \cite[Section~8]{MPS27}, a faster data structure for weighted range LIS, an open problem raised in \cite{Tis08a, GGK25, KR26}. Moreover, even without weights, the complexity of the matrix structure and of the multiplication algorithm has hindered the understanding of problems that otherwise admit straightforward dynamic programs. For instance, only recently was Tiskin's range LIS algorithm shown to support efficient reporting of an optimal subsequence \cite{GGK25,KR26}. A natural alternative is to avoid Monge matrix multiplication altogether. For weighted range LIS, Korogi \cite{Kor26} very recently showed that this is possible, with a data structure based on elementary arguments. So this reduction no longer suggests that the weighted clique problems are hard, and a similar approach might work for them.

\paragraph{Circle trapezoid and polygon-circle graphs.}
Circle trapezoid graphs were introduced by Felsner, M\"uller and Wernisch \cite{FMW97} as a common generalization of trapezoid graphs, circle graphs and circular-arc graphs. They are the intersection graphs of regions between two non-crossing chords of a circle, which we call \emph{c-trapezoids}. The same paper gives an $O(n^2\log n)$-time algorithm for maximum weight clique, which is the best known algorithm for the problem. Polygon-circle graphs, introduced by Koebe as spider graphs \cite{Koe92}, are the intersection graphs of convex polygons inscribed in a circle \cite{KK97}, and as graph classes they contain all the classes above. Recognizing them is NP-complete \cite{Per07}, so the polygons are given as part of the input. To the best of our knowledge, maximum clique has not been studied for this class.

\subsection{Our contribution}

We propose a unifying reduction that, extending the sweep of \cite{MPS27}, turns maximum weight clique in all four classes into the weighted version of their red-blue point problem (Problem~\ref{prob:rb} below). We then give a data structure for this problem that handles weights. It avoids Monge matrices entirely, so even for unit weights it gives simpler algorithms, both for the red-blue point problem and for maximum clique in circle graphs and circular-arc graphs. Combined, the two parts give the following results. Throughout, the input is the list of the $n$ objects and the weights are reals.

\begin{restatable}{theorem}{mainthm}\label{thm:main}
Maximum weight clique in circle trapezoid graphs can be solved in time $O(n\log^2 n)$.
\end{restatable}

This algorithm improves the $O(n^2\log n)$ bound of \cite{FMW97}, both with and without weights. Chords and arcs of the circle are circle trapezoids with models of the same size (Section~\ref{sec:assembl}), so the algorithm also solves the problem for circle and circular-arc graphs.

\begin{restatable}{corollary}{circlecor}\label{cor:circle}
Maximum weight clique in circle graphs can be solved in time $O(n\log^2 n)$.
\end{restatable}

\begin{restatable}{corollary}{arccor}\label{cor:arc}
Maximum weight clique in circular-arc graphs can be solved in time $O(n\log^2 n)$.
\end{restatable}

The same approach also applies to polygon-circle graphs, a larger class that contains all the classes above, where the input size is the total number $N$ of vertices of the polygons. Since a polygon model of a circle trapezoid graph may need many more vertices than the original model has arcs, Theorem~\ref{thm:poly} does not imply Theorem~\ref{thm:main}.

\begin{restatable}{theorem}{polythm}\label{thm:poly}
Maximum weight clique in polygon-circle graphs can be solved in time $O(N\log^2 N)$, where $N$ is the total number of vertices of the polygons.
\end{restatable}

\begin{table}[htb]
\centering\footnotesize
\setlength{\tabcolsep}{3.5pt}
\renewcommand{\arraystretch}{1.3}
\begin{tabular}{@{}lllll@{}}
\toprule
 & \multicolumn{2}{c}{unit weights} & \multicolumn{2}{c}{arbitrary weights} \\
\cmidrule(lr){2-3}\cmidrule(l){4-5}
graph class & previous & this paper & previous & this paper \\
\midrule
circle & $O(n\log^2 n)$ \cite{Tis15} & $O(n\log^2 n)$ & $O(n\log n+\min\{n^2,m\log\log n\})$ \cite{AAH93} & \boldmath$O(n\log^2 n)$ \\
circular-arc & $O(n\log^2 n)$ \cite{MPS27} & $O(n\log^2 n)$ & $O(dn\log\log n+n\log n)$ \cite{MPS27} & \boldmath$O(n\log^2 n)$ \\
circle trapezoid & $O(n^2\log n)$ \cite{FMW97} & \boldmath$O(n\log^2 n)$ & $O(n^2\log n)$ \cite{FMW97} & \boldmath$O(n\log^2 n)$ \\
polygon-circle & --- & \boldmath$O(N\log^2 N)$ & --- & \boldmath$O(N\log^2 N)$ \\
\bottomrule
\end{tabular}
\caption{Maximum clique in the four classes, with $n$ objects, $m$ edges (in the implicit graph representation), minimum load $d\le2m/n$, and $N\ge2n$ polygon vertices. Bold bounds are new or improve on the previous worst-case bounds.}
\label{tab:results}
\end{table}

Table~\ref{tab:results} compares these bounds with previous work. Corollaries~\ref{cor:circle} and~\ref{cor:arc} give the first worst-case subquadratic algorithms for maximum weight clique in circle graphs and in circular-arc graphs, answering the questions of \cite{Tis06} and \cite{MPS27}, respectively.

Each of our algorithms has two parts. The first is a reduction, presented in Sections~\ref{sec:ct} and~\ref{sec:pc}, to the red-blue problem of \cite[Problem~7]{MPS27}, with weights. This problem was introduced for circular-arc graphs, and it turns out to capture the larger classes as well:

\begin{restatable}[Maximum weight compatible bichromatic set, quadrant queries]{problem}{rbrb}\label{prob:rb}
We are given two sets of red and blue points in the plane with $n$ points in total, each with a non-negative weight. For each query point $(X, Y)$, output the maximum total weight of a subset of the points in the quadrant $(-\infty,X]\times(-\infty,Y]$ in which no red point is strictly larger, in both coordinates, than a blue point.
\end{restatable}

The second part, presented in Section \ref{sec:red-blue}, shows how to answer these queries efficiently.

\begin{restatable}{theorem}{rbrbthm}\label{thm:ds}
Problem~\ref{prob:rb} on $n$ points can be preprocessed in $O(n\log^2 n)$ time so that each query is answered in
$O(\log n)$ time.
\end{restatable}

Finally, we note that the data structure presented in Section~\ref{sec:red-blue} can be extended to answer range queries instead of quadrant queries. By combining this extension with \cite[Theorem~24]{MPS27}, we obtain a data structure for weighted range LIS queries, matching the bounds of \cite{Kor26}. We present the proof of Corollary~\ref{cor:dsrange} in Section~\ref{sec:red-blue}.

\begin{restatable}[Maximum weight compatible bichromatic set, interval queries]{problem}{rbrbrange}\label{prob:rbrange}
    We are given two sets of red and blue points in the plane with $n$ points in total, each with a non-negative weight. For each query interval $[L,R]$, output the maximum total weight of a subset of the points whose $y$-coordinates lie in $[L,R]$, in which no red point is strictly larger, in both coordinates, than a blue point.
\end{restatable}

\begin{restatable}{corollary}{rbrbrangethm}\label{cor:dsrange}
    Problem~\ref{prob:rbrange} on $n$ points can be preprocessed in $O(n\log^2 n)$ time so that each query is answered in $O(\log n)$ time. As a result, weighted range LIS also admits a data structure with $O(n\log^2 n)$ preprocessing time and $O(\log n)$ query time. 
\end{restatable}

This does not improve the bounds for weighted range LIS, but it shows that they hold for a more general problem, which may find further applications, geometric or not (see Section~\ref{sec:open}). Our data structure also supports reporting the weighted LIS in $O(k \log^2 n)$ time, where $k$ is the number of elements in the reported subsequence.

\subsection{Technical Overview}

\begin{figure}[htb]
\centering
\input{figs/fig_overview}
\caption{(a)~A chord from the fixed cut point $\varphi$ intersects four of the five circle trapezoids. (b)~The two arcs into which the chord divides the circle, straightened into two parallel lines from $\varphi$ to $\theta$: the c-trapezoids intersected by the chord become trapezoids between them. (c)~In the box representation, the trapezoids become boxes, and a clique is a set of boxes none of which lies above and to the right of another; the staircase passes through the boxes of a maximum clique. (d)~The corners of the boxes, red at the lower left and blue at the upper right, are the points of one fixed set that lie in the quadrant of the chord, plus a free corner at the origin. Moving the chord only moves the apex of the quadrant, along the line $x+y=1$.}
\label{fig:overview}
\end{figure}
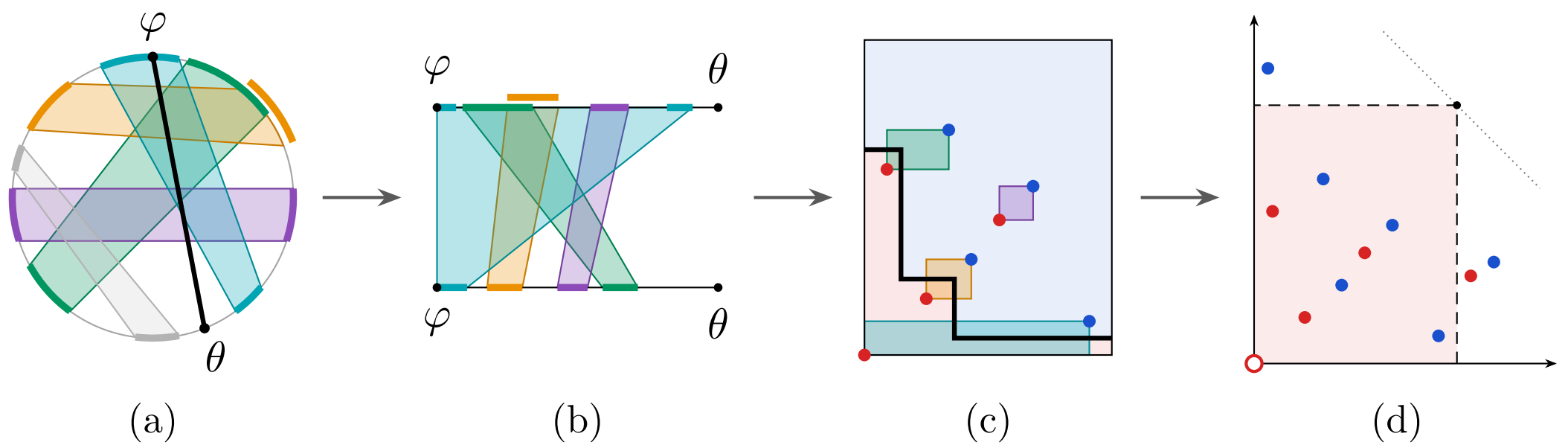

\paragraph{A radial sweep with fixed endpoint.}
Our algorithm for circle trapezoid graphs builds on ideas of Felsner, M\"uller and Wernisch \cite{FMW97}. They show that every clique of c-trapezoids is intersected by one of the chords of its c-trapezoids, and that the c-trapezoids intersected by a chord form a trapezoid graph, in which a maximum weight clique can be found in $O(n\log n)$ time. Trying all $2n$ chords of the c-trapezoids gives $O(n^2\log n)$ time, but the $2n$ instances are unrelated, so no work is shared between them. We also use a chord intersecting the clique, but we choose it so that the instances are related. For arcs, one endpoint of the chord can be fixed once and for all \cite{MPS27}: for an arbitrary cut point $\varphi$ and every clique, some chord $\varphi\theta$ intersects all arcs of the clique. Our first observation is that the same holds for c-trapezoids and polygons. Cut the circle at $\varphi$. The chord $\varphi\theta$ intersects an object exactly when $\theta$ lies in its \emph{span}, the smallest interval of the cut circle containing it, and two intersecting objects have intersecting spans. By the Helly property of intervals (pairwise intersecting intervals share a point), all spans of a clique share a point $\theta$ (Fig.~\ref{fig:overview}a). Hence it suffices to sweep $\theta$ around the circle and compute, for every $\theta$, a maximum weight clique among the objects intersected by $\varphi\theta$: the best of these is a maximum weight clique. 

\paragraph{From chords to boxes.}
The chord $\varphi\theta$ splits the circle into two arcs, which, as observed by Felsner, M\"uller and Wernisch \cite{FMW97}, act as the two parallel lines of a trapezoid graph \cite{DGP88} (Fig.~\ref{fig:overview}b). Each object intersected by the chord has points on both arcs, and two such objects are disjoint exactly when, on both arcs, all points of one of them lie between $\varphi$ and all points of the other. Through their box representation of trapezoid graphs, every such object becomes an axis-parallel box, and a clique becomes a set of boxes none of which lies strictly above and to the right of another (Fig.~\ref{fig:overview}c).

\paragraph{From boxes to red and blue points.}
Consider a \emph{staircase}, a path of horizontal and vertical segments descending from left to right. Since it is monotone, it never passes through two boxes one of which lies strictly above and to the right of the other, so the boxes it passes through form a clique; conversely, the boxes of every clique are passed through by one staircase (Fig.~\ref{fig:overview}c).  So we look for a staircase that passes through boxes of maximum total weight. To turn boxes into points, put a red point at the lower-left corner and a blue point at the upper-right corner of every box, both with the weight of the object, and say that a staircase \emph{scores} the red points on or below it and the blue points on or above it. A staircase scores both corners of every box it passes through and exactly one corner of every other box, so a staircase of maximum score on the points also passes through boxes of maximum total weight. A set of points is scored by one staircase exactly when no red point lies strictly above and to the right of a blue point, which is the condition of Problem~\ref{prob:rb}. This is the argument of \cite[Theorem~24]{MPS27}, with boxes instead of points.

\paragraph{One point set for all the chords.}
The key observation is that the corners depend on $\theta$ only in a very limited way, so we can build a single set of points shared by all the chords. Measure positions from $\varphi$ along both arcs. The lower-left corner of a box is where its object starts, so it does not depend on $\theta$ at all. The upper-right corner does depend on $\theta$, but in a controlled way, and with a little more care (Section~\ref{sec:ct}) it can also be taken from a fixed set of points. So the corners for the chord $\varphi\theta$ are the points of one fixed set inside a quadrant whose apex slides along the line $x+y=1$ as $\theta$ moves (Fig.~\ref{fig:overview}d), apart from a few corners at the origin and at the apex, which are easy to handle. This is where the fixed endpoint pays off: the $2n$ unrelated trapezoid-graph instances of Felsner, M\"uller and Wernisch become queries to a single data structure.

\paragraph{Weighted quadrant queries without Monge matrix multiplication.}
To tackle Problem~\ref{prob:rb}, we develop a new algorithm that replaces the previous algorithm of~\cite{MPS27} for the unweighted case. We devise a dynamic program for maximum-scoring staircases with $\Theta(n^3)$ entries, and show how to compute all these entries implicitly in $O(n\log^2 n)$ time. By implicitly, we mean that our algorithm provides efficient oracle access to every entry of the DP table in $O(\operatorname{polylog} n)$ time, which is powerful enough to solve not only Problem~\ref{prob:rb}, but also the related Problem~\ref{prob:rbrange}. Our algorithm avoids core-sparse Monge matrix multiplication and relies entirely on dynamic programming and binary tree manipulations. This brings three benefits: (1) the algorithm uses only elementary technical tools, and the objects it computes admit simple definitions; (2) consequently, it admits a straightforward reporting algorithm for optimal staircases and weighted LIS; and (3) it runs in $O(n\log^2 n)$ time, only an $O(\log n)$ factor away from the lower bound. Achieving comparable bounds through the core-sparse approach of~\cite{MPS27} would require proving an $O(n)$ bound on the core size for weighted LIS, which currently seems out of reach, if such a bound even holds.

\paragraph{More details on our weighted quadrant queries.}
The transitions in our DP consist of adding a constant to a range of rows and replacing each row in $[0,a-1]$ by its pointwise maximum with row $a$, requiring $O(n^2)$ time per element naively. We use two ideas to replace each of the two $O(n)$ factors with an $O(\log n)$ factor, culminating in $O(\log^2 n)$ time per element. The first is a monotonicity property derived from the planarity of the underlying paths, which allows us to compute the pointwise maximum of two rows by binary search and concatenation of two binary trees in $O(\log n)$ time. The second is to apply these maximum operations to a range of rows using lazy propagation, while carefully maintaining the structural properties needed for the monotonicity argument. The idea of exploiting monotonicity to compute pointwise maxima originates from~\cite{Kor26}, but our setting is more challenging because we maintain a much larger DP table and apply pointwise maxima to entire ranges of rows. Lazy propagation introduces intermediate vectors that may be outdated and need not directly correspond to rows of the current DP table. We overcome this difficulty by proving that the pointwise maxima computed during lazy propagation always produce valid rows of the DP table, allowing us to exploit the monotonicity property even in this setting.

\subsection{Open problems}\label{sec:open}

\paragraph{Making the algorithm optimal.}
Our algorithms take $O(n\log^2 n)$ time, and a natural question is whether $O(n\log n)$ time is possible, both with and without weights. An $O(n\log n)$ bound would be the best possible for comparison-based algorithms: deciding whether $n$ numbers are pairwise distinct takes $\Omega(n\log n)$ comparisons \cite{BenOr83}, and easily reduces to deciding whether the clique number is at least~$2$.

\paragraph{Other reductions to the red-blue point problem.}
The red-blue point problem turns out to be quite expressive. It was introduced for circular-arc graphs \cite{MPS27} and the same paper reduces weighted range LIS to its variant with interval queries. We show that it also captures maximum weight clique in circle trapezoid graphs and polygon-circle graphs, which are much more general classes than circular-arc graphs. It would be interesting to understand whether other problems, especially geometric ones, reduce to it or to its variant with intervals.

\paragraph{Improvements for problems dependent on Monge matrix multiplication.}
Fast multiplication of unit-Monge and core-sparse Monge matrices has achieved great success in obtaining highly efficient algorithms for problems with sparse Monge structure, but its complexity and difficulty in generalizing to weighted problems have posed obstacles to further progress. We believe that our approach, together with~\cite{Kor26}, demonstrates new techniques that could be applied to these problems. An interesting direction for future research is to explore whether these techniques can be extended to yield improvement for other problems that currently rely on Monge matrix multiplication.

\section{Preliminaries}\label{sec:prelim}
\paragraph{Weights.}
We assume without loss of generality that all weights are non-negative: an object of negative weight is never part of a maximum weight clique, as removing it from a clique leaves a clique of larger weight, so such objects can be discarded in advance.

\paragraph{Red and blue points.}
We use the terminology of \cite{MPS27}. A red point \emph{dominates} a blue point if it is strictly larger in both coordinates, and a set of red and blue points is \emph{compatible} if none of its red points dominates one of its blue points. For a finite set $X$ of points with non-negative weights, $\mwcs(X)$ denotes the maximum weight of a compatible subset of $X$; Problem~\ref{prob:rb} asks for $\mwcs(X\cap Q)$ for quadrants $Q$.

A red point and a blue point with a common coordinate never dominate each other. So we may sort the points by $x$-coordinate, putting red points before blue points with the same $x$-coordinate, do the same for $y$, and replace every coordinate by its rank. This neither changes which red points dominate which blue points nor which points lie in a quadrant
\cite[Lemma~14]{MPS27}. 
For Problem~\ref{prob:rb}, we further assume that each point coordinate is a half-integer in $\{\frac{1}{2}, 1+\frac{1}{2}, \ldots, n-\frac{1}{2}\}$, and that each query point $(X,Y)$ has integer coordinates in $[0,n]$. The second assumption can be achieved by computing the rank of each query coordinate in $O(\log n)$ time using binary search.

A \emph{staircase} is a non-increasing function $\Sigma:\mathbb R\to\mathbb R\cup\{\pm\infty\}$. It \emph{scores} the red points below it and the blue points above it. We only consider staircases that do not pass through any points. 

\begin{lemma}[{\cite[Lemma~15]{MPS27}}]\label{lem:stair}
The points scored by a staircase are compatible, and every compatible set is scored by some staircase.
\end{lemma}

The first half is a one-line argument that we use below: if a scored red point $r$ and a scored blue point $b$ have $x_r>x_b$, then $y_r\le\Sigma(x_r)\le\Sigma(x_b)\le y_b$.

\paragraph{Data Structures.} We employ a \emph{segment tree} \cite{LS13} (also known as a \emph{tournament tree}) as the underlying data structure for our dynamic programming implementation. Briefly, a segment tree is a static, balanced binary tree whose leaves correspond to $n$ fixed elements. Each of its $n-1$ internal nodes stores the range aggregate (formally, a monoid product) of the elements in its subtree. Since the tree is static, it does not support the insertion or deletion of elements. Instead, it is initialized as a vector of $n$ elements and supports updates to existing elements. The data structure is simple and especially useful for us because we require all trees to have a fixed internal layout (specifically, for binary searches and concatenations involving two segment trees), whereas self-balancing binary search trees change their layouts dynamically. In Table~\ref{tab:segtree}, we list the operations supported by the segment tree representing a single length-$n$ vector. These operations can be implemented by augmenting the tree with subtree
maxima as a range aggregate, and additional lazy tags for subtree updates. 

Note that our segment tree is fully persistent~\cite{DSST89}. Hence, all operations are immutable, meaning that the input vectors are neither modified nor destroyed during any operation.

\begin{table}[th]
\centering
\renewcommand{\arraystretch}{1.15}
\begin{tabularx}{\textwidth}{lXl}
\hline
Operation & Description & Time \\
\hline
Initialization &
Given a value $x$, return a vector whose entries are all initialized to $x$. &
$O(n)$ \\

Point Update &
Given a vector $v$, a position $1 \le p \le n$, and a value $x$, return a new vector obtained by setting $v_p = x$. &
$O(\log n)$ \\

Range Addition &
Given a vector $v$, a constant $x$, and a range $1 \le \ell \le r \le n$, return a new vector $(v_1, \ldots, v_{\ell-1}, \textcolor{blue}{v_\ell + x, v_{\ell+1} + x, \ldots, v_r + x}, v_{r+1}, \ldots, v_n)$. &
$O(\log n)$ \\

Range Maximum &
Given a vector $v$ and a range $1 \le \ell \le r \le n$, return $\max_{\ell \le i \le r} v_i$. &
$O(\log n)$ \\

Binary Search &
Given two vectors $v,w$ and a monotone Boolean predicate $f$, find the last $j \in [n]$ such that $f(v_j,w_j) = \textsf{FALSE}$, or return $0$ if no such index exists. &
$O(\log n)$ \\

Concatenation &
Given two vectors $v,w$ and a position $0 \le j \le n$, return a new vector $(v_1,v_2,\ldots,v_j,w_{j+1},\ldots,w_n)$. &
$O(\log n)$ \\
\hline
\end{tabularx}
\caption{Operations supported by the segment tree. All vectors have length $n$.}
\label{tab:segtree}
\end{table}

We say that a Boolean predicate $f:\mathbb{Z}\to\{\mathsf{TRUE},\mathsf{FALSE}\}$ is monotone if $f(i)\implies f(i+1)$ for every $i\in\mathbb{Z}$. In this paper, we consider only simple predicates that can be evaluated in $O(1)$ time.

We use the following lemma:
\begin{lemma}[{\cite[Lemma~4]{ABR00}}]\label{lem:twodimrangequery}
Given $n$ two-dimensional points with real weights, we can preprocess them in $O(n \log n)$ time so that the sum of the weights of the points within any axis-parallel rectangle can be computed in $O(\log n)$ time.
\end{lemma}

\section{The reduction for circle trapezoid graphs}\label{sec:ct}

\begin{figure}[htb]
\centering
\input{figs/ct_fig_model.tex}
\caption{(a)~An instance with five circle trapezoids. The numbers mark their arcs. (b)~Lemma~\ref{lem:gap} for $T_2$ (orange). Its gaps $G$ and $G'$ are dotted, and the regions $R_G$ and $R_{G'}$ between them and $T_2$ are shaded. Both arcs of $T_3$ lie in $G$, so $T_3$ lies in $R_G$ and does not intersect $T_2$. An arc of $T_1$ overlaps an arc of $T_2$, so it lies in no gap of $T_2$, and $T_1$ intersects $T_2$.}
\label{fig:ctmodel}
\end{figure}
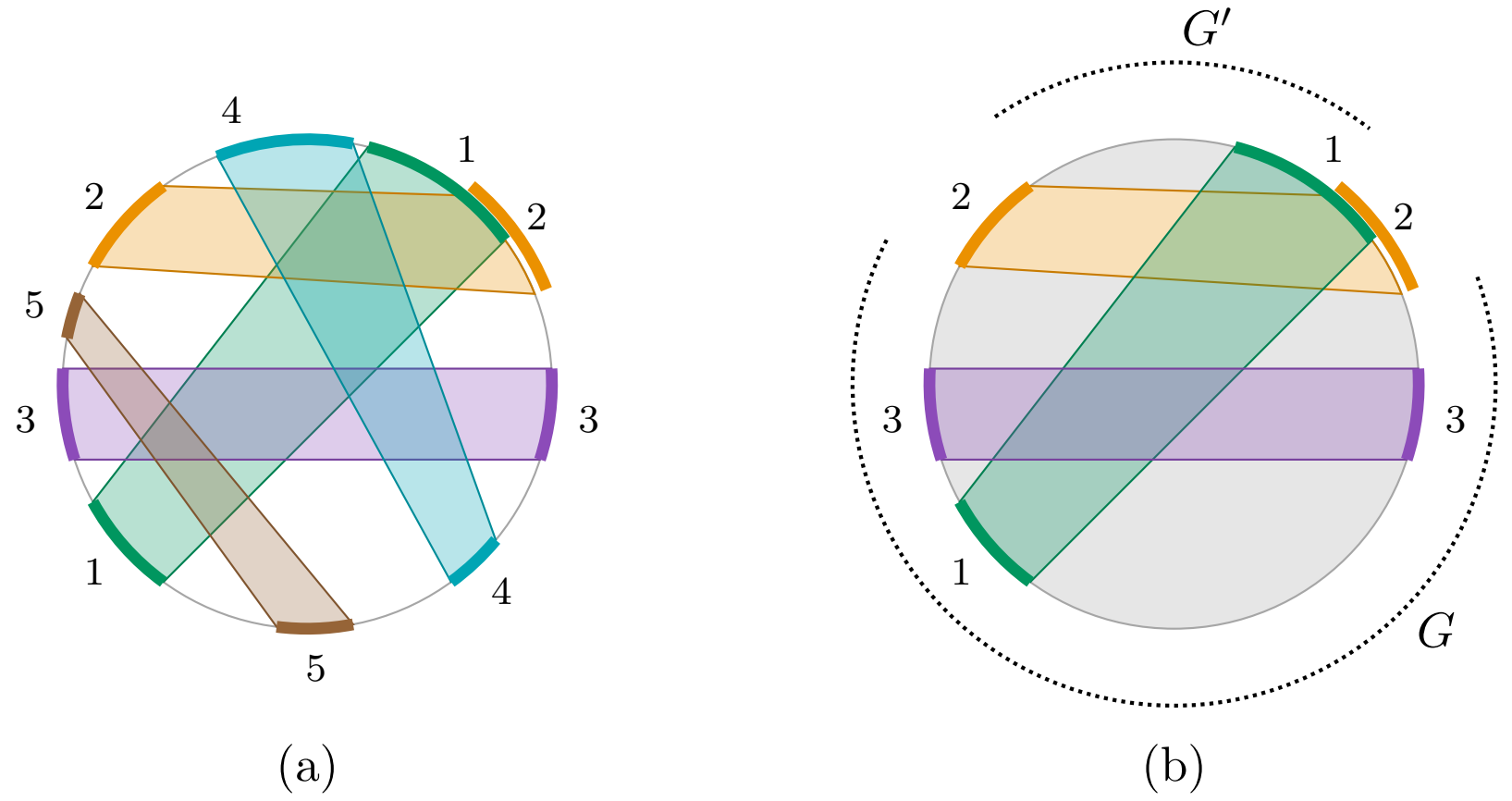

Throughout, $C$ is a circle, and the \emph{disk} is the closed region it bounds. A \emph{circle trapezoid}, or \emph{c-trapezoid} for short, is the convex hull $T=\operatorname{conv}(A\cup B)$ of two disjoint closed arcs $A$ and $B$ of $C$
\cite{FMW97}: if $A$ runs clockwise from $a$ to $a'$ and $B$ from $b$ to $b'$, then $T$ is the region bounded by $A$, the chord $a'b$, $B$ and the chord $b'a$. Equivalently, it is the region of the disk between two non-crossing chords (Fig.~\ref{fig:ctmodel}a). Removing $A$ and $B$ from $C$ leaves two open arcs, the \emph{gaps} of $T$ (Fig.~\ref{fig:ctmodel}b).

We are given $n$ c-trapezoids, each with a weight $w(T)\ge0$, and look for a \emph{maximum weight clique}, a set of pairwise intersecting c-trapezoids of maximum total weight. As usual, we assume without loss of generality that the $4n$ endpoints of the arcs are pairwise distinct.

\begin{lemma}\label{lem:gap}
Two c-trapezoids $T$ and $T'$ are disjoint if and only if both arcs of $T'$ lie entirely in the same gap of $T$.
\end{lemma}

\begin{proof}
For a gap $G$ of $T$, let $R_G$ be the region between $G$ and $T$: the part of the disk on the side of the chord of $T$ joining the ends of $G$, without the chord itself (Fig.~\ref{fig:ctmodel}b shows the two regions for $T=T_2$). The two regions are convex and disjoint, their union is the part of the disk outside $T$, and $R_G$ intersects the circle exactly in $G$. If both arcs of $T'$ lie entirely in a gap $G$ of $T$, they lie in $R_G$, and so does their convex hull $T'$, which therefore does not intersect $T$. Conversely, if $T'$ does not intersect $T$, then $T'$ lies in the disk outside $T$. Specifically, it lies in a single region $R_G$, so its arcs lie in $G$.
\end{proof}

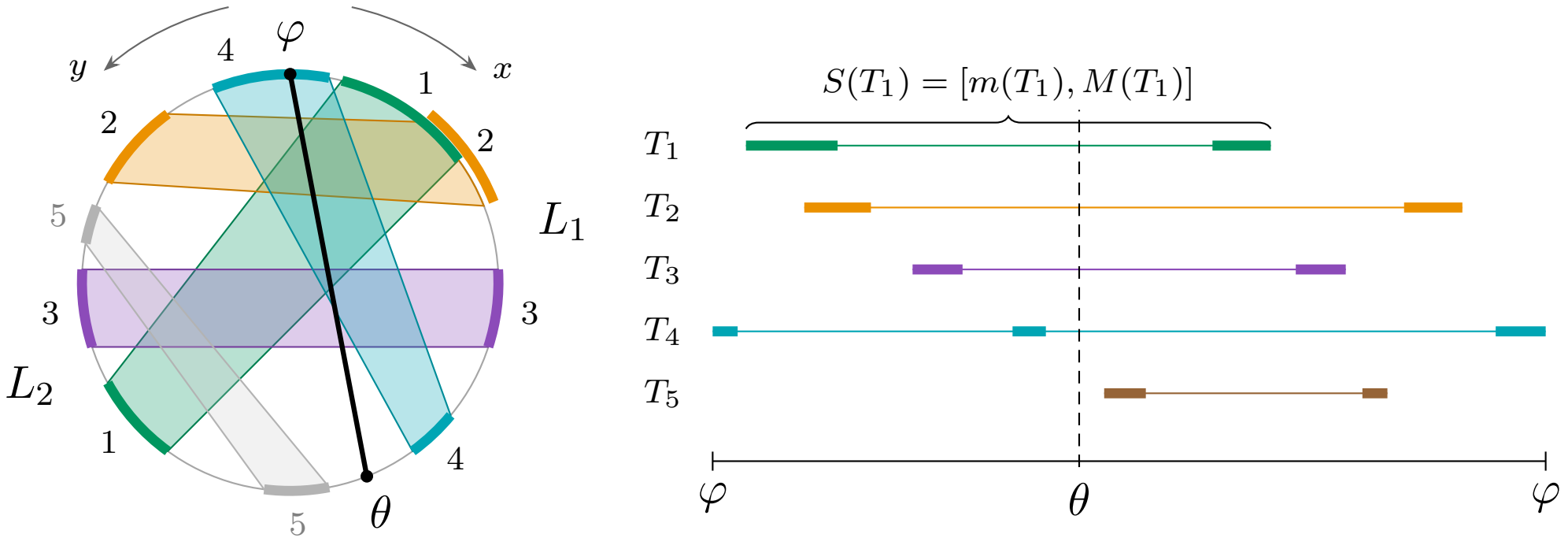
\begin{figure}[htb]
\centering
\input{figs/ct_fig_chord.tex}
\caption{(a)~The chord $\varphi\theta$ intersects $T_1,\dots,T_4$, among them the clique $\{T_1,T_2,T_4\}$, and does not intersect $T_5$ (grey). It divides the circle into the arcs $L_1$ and $L_2$, and the arrows show the coordinates $x$ and $y$, which grow from $\varphi$ to $\theta$. (b)~The circle cut open at $\varphi$. Each row shows the arcs of a c-trapezoid (thick) and its span (thin). $T_4$ contains $\varphi$, so its first arc appears at both ends and its span is the whole line. The chord intersects exactly the c-trapezoids whose span crosses the dashed line.}
\label{fig:ctchord}
\end{figure}

Following Moschetti, Polak and Schmidt \cite{MPS27}, we fix a point $\varphi$ of $C$ that is not an endpoint of an arc, the \emph{cut point}, and consider the chords that start at $\varphi$. We show that for every clique one of these chords intersects all its c-trapezoids (Lemma~\ref{lem:chord}), so it suffices to look, one chord at a time, at the c-trapezoids intersected by the chord. To describe them, we cut the circle open at $\varphi$: we measure positions on $C$ clockwise from $\varphi$, so that $C\setminus\{\varphi\}$ becomes the open interval $(0,1)$, and for $\theta\in(0,1)$ the \emph{chord} $\varphi\theta$ is the segment from $\varphi$ to the point at position $\theta$ (Fig.~\ref{fig:ctchord}a). From now on, \emph{first}, \emph{last},
\emph{before} and \emph{after} refer to this order of positions.

\paragraph{Spans.}
Suppose first that $\varphi$ lies in no arc of $T$. Let $m(T)$ and $M(T)$ be the first and the last point of $A\cup B$, that is, the smallest and the largest position of a point of $A\cup B$. The \emph{span} $S(T)=[m(T),M(T)]$ is the smallest interval of $(0,1)$ containing both arcs, that is, the smallest arc of $C$ that contains both arcs and avoids $\varphi$ (the thin lines in Fig.~\ref{fig:ctchord}b). One gap of $T$ is the rest of the circle, around $\varphi$; the other lies inside the span, between the two arcs, and we call it the \emph{inner gap}. If $\varphi$ lies in an arc of $T$, we set $S(T)=(0,1)$; then both gaps of $T$ avoid $\varphi$. We only use values of $\theta$ that are not endpoints.

\begin{lemma}\label{lem:span}
The chord $\varphi\theta$ intersects $T$ if and only if $\theta\in S(T)$.
\end{lemma}

\begin{proof}
The chord is a c-trapezoid whose two arcs are the points $\varphi$ and $\theta$. By Lemma~\ref{lem:gap}, it does not intersect $T$
if and only if $\varphi$ and $\theta$ lie in the same gap of $T$. If $\varphi$ lies in an arc of $T$ this never happens.
Otherwise the gap containing $\varphi$ is the one outside the span, and it contains $\theta$ exactly when $\theta\notin S(T)$.
\end{proof}

\begin{lemma}\label{lem:chord}
For every clique $K$ there is a chord $\varphi\theta$ that intersects all c-trapezoids of $K$.
\end{lemma}

\begin{proof}
Two c-trapezoids with disjoint spans are disjoint: the arcs of one lie in its span, so they lie outside the span of the other, in its gap around $\varphi$, and Lemma~\ref{lem:gap} applies. The spans of the c-trapezoids of $K$ therefore pairwise intersect, so by the Helly property of intervals they have a common part. This common part is an interval whose two ends are distinct endpoints of arcs (or $0$ and $1$), so it contains a point $\theta$ that is not an endpoint. By Lemma~\ref{lem:span}, the chord $\varphi\theta$ intersects all c-trapezoids of $K$.
\end{proof}

By Lemma~\ref{lem:span}, the c-trapezoids intersected by the chord depend only on which endpoints lie before $\theta$, so it suffices to try one
$\theta$ in each of the $4n+1$ intervals into which the endpoints cut $(0,1)$.\footnote{A more careful analysis shows that $n+1$ chords suffice: $\theta$ can always be taken in one of the at most $n+1$ of these intervals that start at $0$ or at some $m(T)$.}

\medskip
From here on, we apply the method of Felsner, M\"uller and Wernisch \cite{FMW97} to one chord $\varphi\theta$ at a time: the c-trapezoids it intersects form a trapezoid graph, which we handle through their box representation. Unlike them, we use the chords from $\varphi$ instead of the $2n$ chords of the c-trapezoids, so that most of the structure is preserved from one chord to the next, as outlined in the overview.

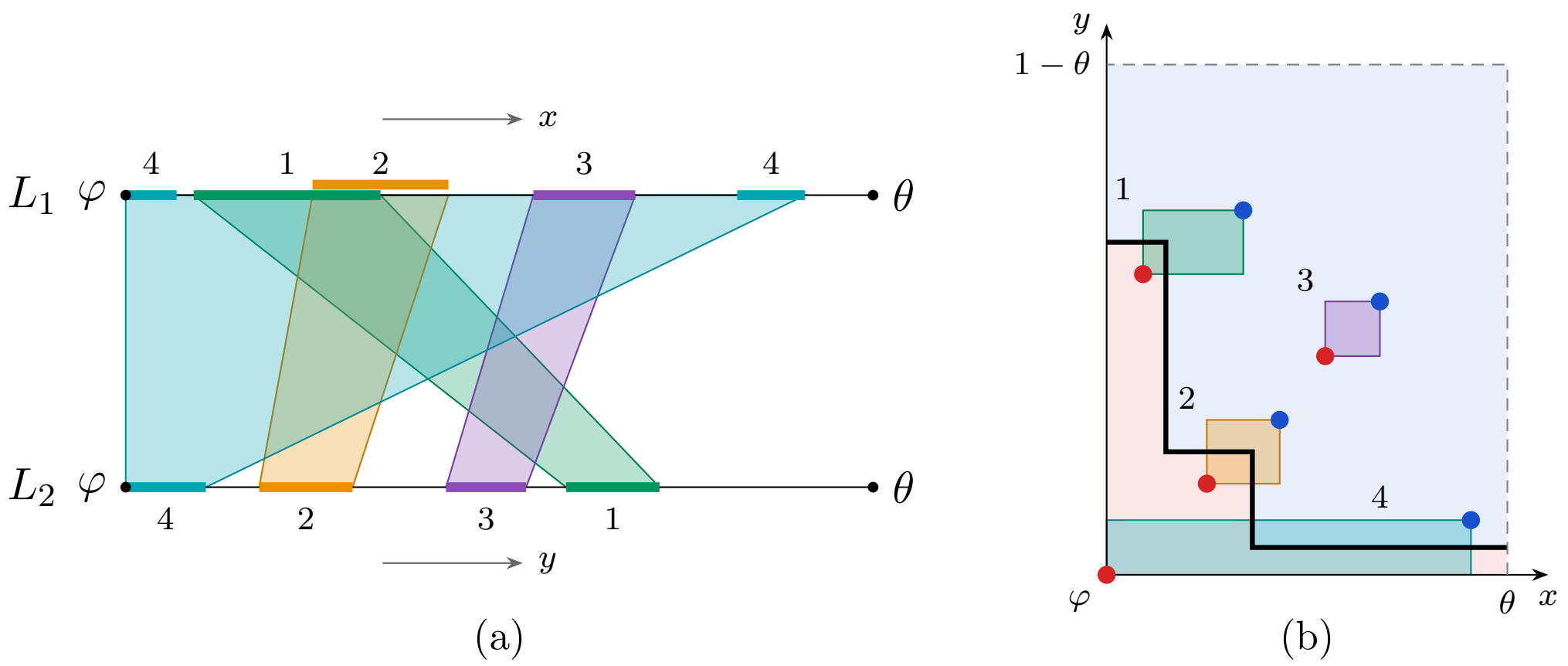
\begin{figure}[htb]
\centering
\input{figs/ct_fig_lines.tex}
\caption{(a)~The arcs $L_1$ and $L_2$ from $\varphi$ to $\theta$ into which the chord $\varphi\theta$ divides the circle, drawn as two parallel lines, and each c-trapezoid intersected by the chord, drawn as a quadrilateral between them with its arcs thick. $T_2$ lies to the left of $T_3$ on both lines, so they are disjoint; the arcs of $T_1$ and $T_2$ overlap on $L_1$. (b)~Their boxes $[x_1,x_2]\times[y_1,y_2]$, with red lower-left and blue upper-right corners. The staircase passes through the boxes of the clique $\{T_1,T_2,T_4\}$: their red corners lie on or below it and their blue corners on or above it. Of $T_3$ it scores only the blue corner. Neither of the boxes of $T_1$ and $T_2$ dominates the other, because their $x$-ranges overlap, although the boxes themselves are disjoint.}
\label{fig:ctlines}
\end{figure}

Fix $\theta$, and let $\T_\theta$ be the set of c-trapezoids intersected by the chord $\varphi\theta$. The chord divides $C$ into two arcs from $\varphi$ to $\theta$, which share their ends: $L_1$, clockwise from $\varphi$, and $L_2$, counterclockwise from $\varphi$ (Fig.~\ref{fig:ctchord}a). On $L_1$ we use the coordinate $x$, the distance from $\varphi$ clockwise, and on $L_2$ the coordinate $y$, the distance from $\varphi$ counterclockwise. So a point at position $t$ has $x=t$ if it lies on $L_1$ and $y=1-t$ if it lies on $L_2$, the point $\varphi$ has $x=y=0$, and $\theta$ has $x=\theta$ and $y=1-\theta$; both coordinates grow from $\varphi$ to $\theta$. Every $T\in\T_\theta$ has points on both $L_1$ and $L_2$: an arc through $\varphi$ reaches both, and otherwise $m(T)<\theta<M(T)$. Let $x_1$ and $x_2$ be the smallest and the largest $x$-coordinate of a point of $A\cup B$ on $L_1$, and $y_1$ and $y_2$ the same on $L_2$ (Fig.~\ref{fig:ctlines}a).

\begin{observation}\label{obs:ends}
Let $T\in\T_\theta$. If $\varphi$ lies in an arc of $T$, then $x_1=y_1=0$; otherwise $x_1=m(T)$ and $y_1=1-M(T)$.
If $\theta$ lies in an arc of $T$, then $x_2=\theta$ and $y_2=1-\theta$; otherwise $\theta$ lies in a gap $(p,q)$ of $T$, and
$x_2=p$ and $y_2=1-q$.
\end{observation}

Draw $L_1$ and $L_2$ as two parallel lines, both going from $\varphi$ on the left to $\theta$ on the right, and replace each $T\in\T_\theta$ by the quadrilateral with side $[x_1,x_2]$ on $L_1$ and side $[y_1,y_2]$ on $L_2$ (Fig.~\ref{fig:ctlines}a). This is the usual picture of a trapezoid graph \cite{DGP88}, in which two quadrilaterals are disjoint exactly when one of them lies to the left of the other on both lines. Felsner, M\"uller and Wernisch show that the c-trapezoids intersected by a chord of a c-trapezoid form a trapezoid graph in this way \cite{FMW97}; the next lemma shows the same for the chords $\varphi\theta$.

\begin{lemma}\label{lem:lines}
Two c-trapezoids $T,T'\in\T_\theta$ are disjoint if and only if $x_2(T')<x_1(T)$ and $y_2(T')<y_1(T)$, or the same
holds with $T$ and $T'$ exchanged.
\end{lemma}
\begin{proof}
Suppose that $T$ and $T'$ are disjoint. By Lemma~\ref{lem:gap}, the arcs of $T'$ lie in one gap $G$ of $T$. The arc
$G$ contains points of $T'$ on both $L_1$ and $L_2$. So $G$ contains exactly one of $\varphi$ and $\theta$, the two points where $L_1$ and $L_2$ meet. If $G$ contains $\varphi$, then $G$ consists of the points with $x<a$ on $L_1$ and the points with $y<b$ on $L_2$, for some
$a$ and $b$. Since $T$ has no point in $G$, $x_1(T)\ge a$ and $y_1(T)\ge b$. The point of $T'$ with $x$-coordinate $x_2(T')$ on $L_1$ and the point with $y$-coordinate
$y_2(T')$ on $L_2$ lie in $G$, so $x_2(T')<a$ and $y_2(T')<b$. Hence $x_2(T')<x_1(T)$ and $y_2(T')<y_1(T)$. If $G$
contains $\theta$, then $G$ consists of the points with $x>a$ on $L_1$ and the points with $y>b$ on $L_2$, for some $a$
and $b$, and in the same way $x_2(T)\le a<x_1(T')$ and $y_2(T)\le b<y_1(T')$.

Conversely, let $x_2(T')<x_1(T)$ and $y_2(T')<y_1(T)$. The points with $x<x_1(T)$ on $L_1$ and the points with
$y<y_1(T)$ on $L_2$ form an open arc of $C$, which contains $\varphi$ because $x_1(T)>x_2(T')\ge0$ and
$y_1(T)>y_2(T')\ge0$. This arc contains every point of the arcs of $T'$ and no point of $T$, so it lies in a single gap
of $T$, and $T$ and $T'$ are disjoint by Lemma~\ref{lem:gap}.
\end{proof}

A \emph{box} is a set $[x_1,x_2]\times[y_1,y_2]$. A box $b$ \emph{dominates} a box $b'$ if it lies strictly above and to the right of $b'$, that is, if $x_1(b)>x_2(b')$ and $y_1(b)>y_2(b')$. Give every $T\in\T_\theta$ the box $[x_1,x_2]\times[y_1,y_2]$ (Fig.~\ref{fig:ctlines}b). Lemma~\ref{lem:lines} says that two c-trapezoids of $\T_\theta$ are disjoint if and only if one of their boxes dominates the other, so a set of c-trapezoids of $\T_\theta$ is a clique if and only if none of their boxes dominates another.

We say that a staircase \emph{passes through} a box if the lower-left corner of the box lies on or below the staircase and its upper-right corner lies on or above it (Fig.~\ref{fig:ctlines}b). Since a staircase goes down from left to right, it cannot pass through two boxes one of which dominates the other. Conversely, we show below that every set of boxes none of which dominates another is passed through by one staircase. So a maximum weight clique of $\T_\theta$ corresponds to a staircase that passes through boxes of maximum
total weight.

This can be formalized as a problem on red and blue points. Put a red point at the lower-left corner and a blue point at the upper-right corner of every box, both with the weight of the box; these are its \emph{corners}. A box dominates another exactly when its red corner dominates the blue corner of the other, which is why we use the same word. A staircase passes through a box exactly when it scores both of its corners, and otherwise it scores one of them. So a staircase scores the total weight $W$ of the boxes plus the weight of the boxes it passes through, which gives the following lemma. It is the argument of \cite[Theorem~24]{MPS27}, with boxes in place of points.

\begin{lemma}\label{lem:boxes}
Consider a family of boxes of total weight $W$, in which the red corner of a box never shares a coordinate with the blue corner of another box. The maximum weight of a set of boxes none of which dominates another equals $\mwcs(\text{corners})-W$. Moreover, the boxes with both corners in a maximum weight compatible set of corners form such a set of maximum weight.
\end{lemma}

\begin{proof}
We may assume without loss of generality that no two corners share a coordinate (otherwise, it suffices to replace the coordinates by their ranks as in Section~\ref{sec:prelim}).

Let $I$ be a set of boxes none of which dominates another. The corners of the boxes of $I$ are compatible: a red corner does not dominate the blue corner of another box of $I$, since neither box dominates the other, nor the blue corner of its own box. By Lemma~\ref{lem:stair}, some staircase $\Sigma$ scores all of them, that is, passes through every box of $I$. It also scores a corner of every other box: if it does not score the red corner $(x_1,y_1)$ of a box, then $\Sigma(x_2)\le\Sigma(x_1)<y_1\le y_2$, and it scores the blue corner $(x_2,y_2)$. The corners scored by $\Sigma$ are compatible (Lemma~\ref{lem:stair}) and weigh at least $2w(I)+(W-w(I))=W+w(I)$, so $\mwcs(\text{corners})\ge W+w(I)$.

Conversely, let $X$ be a compatible set of corners, and $I$ the set of boxes with both corners in $X$. No box of $I$ dominates another, since $X$ is compatible. Every other box has at most one corner in $X$, so $w(X)\le 2w(I)+(W-w(I))=W+w(I)$. Both inequalities together give the equality and the last claim.
\end{proof}

Giving every box of $\T_\theta$ the weight of its c-trapezoid and writing $W_\theta$ for the total weight of $\T_\theta$, Lemmas~\ref{lem:lines} and~\ref{lem:boxes} give
\begin{equation}\label{eq:theta}
\text{maximum weight of a clique of } \T_\theta \;=\; \mwcs(\text{corners of the boxes of } \T_\theta)-W_\theta .
\end{equation}

\begin{figure}[htb]
\centering
\input{figs/ct_fig_points.tex}
\caption{The point set $\Pi$: every c-trapezoid has the red point of its span and the blue point of its inner gap, except $T_4$, which contains $\varphi$ and has the blue points of its two gaps instead. (a)~The quadrant $Q_\theta$ for the chord of Figs.~\ref{fig:ctchord} and~\ref{fig:ctlines}; its points are the corners of Fig.~\ref{fig:ctlines}b, except the free red corner $(0,0)$ of $T_4$. (b)~A second chord, with $\theta'$ inside the first arc of $T_5$: the quadrant now contains the red point of $T_5$, and the blue corner of $T_5$ is the free apex. As $\theta$ moves, only the apex moves, along the line $x+y=1$.}
\label{fig:ctpoints}
\end{figure}
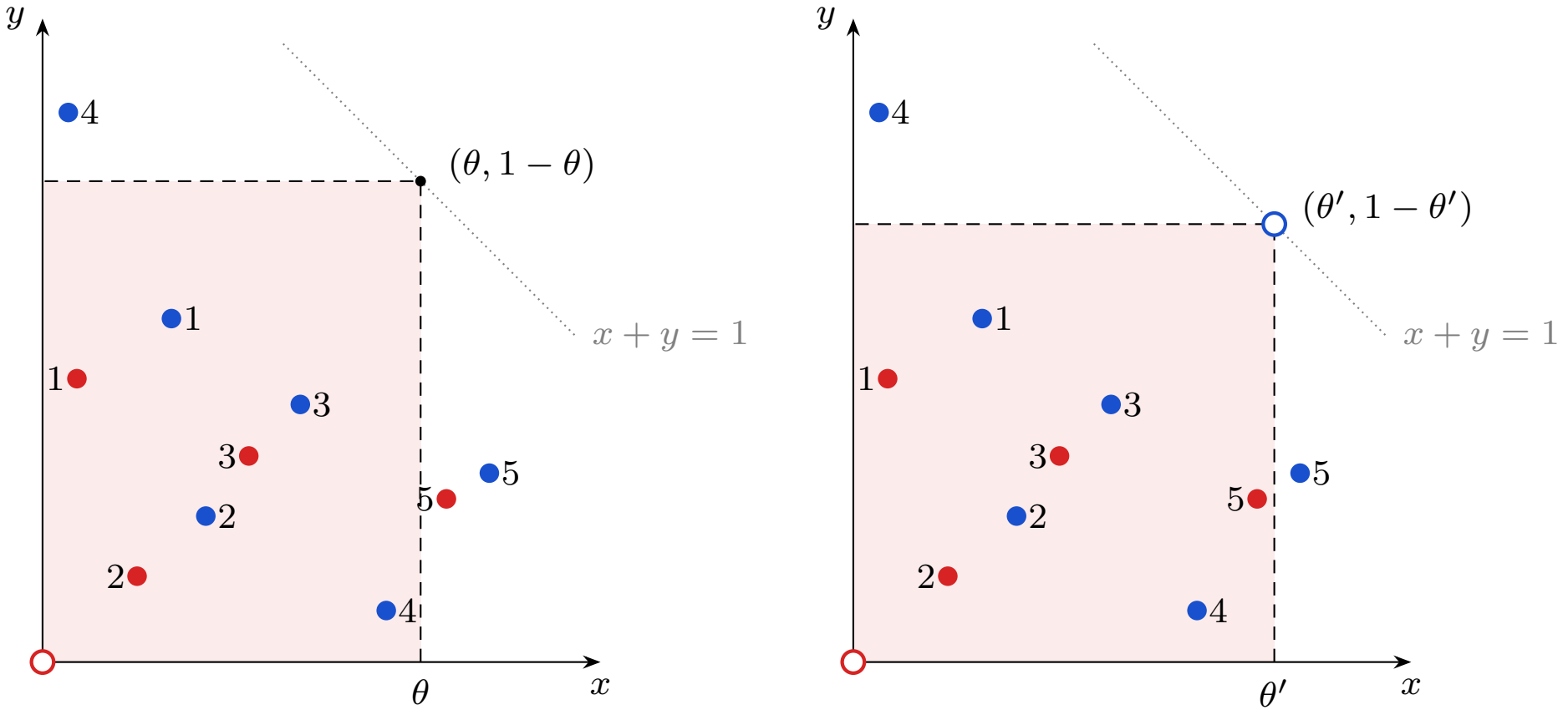

\paragraph{One point set for all chords.}
Up to here, every chord gives its own trapezoid-graph instance, as in \cite{FMW97}, and solving the instances one by one would take quadratic time. Choosing all chords through the cut point changes this: the corners of all instances come from one fixed point set, so that all instances become queries on a single data structure. For an interval $I\subseteq[0,1]$ with endpoints $a<b$ we let $\pi(I)=(a,1-b)$, and for $\theta\in(0,1)$ we let $Q_\theta=(-\infty,\theta]\times(-\infty,1-\theta]$ be the quadrant with apex $(\theta,1-\theta)$. For every $\theta$ that is not an endpoint of $I$,
\begin{equation}\label{eq:pi}
\pi(I)\in Q_\theta \iff a\le\theta\le b \iff \theta\in I .
\end{equation}
In this notation, Observation~\ref{obs:ends} says that the box of $T\in\T_\theta$ has red corner $\pi(S(T))$, which is $(0,0)$ when $\varphi$ lies in an arc of $T$, and blue corner $\pi(G)$ for the gap $G$ of $T$ containing $\theta$, or the apex $(\theta,1-\theta)$ of $Q_\theta$ when $\theta$ lies in an arc of $T$. So the red corner does not depend on $\theta$ at all, and the blue corner depends on $\theta$ only through the gap containing it. We build, once and for all, the set $\Pi$ of the points used across all the instances, each with the weight of its c-trapezoid. For each $\theta$, the quadrant $Q_\theta$ will select the corners for the chord $\varphi\theta$ (Lemma~\ref{lem:quadrants}). If $\varphi$ lies in no arc of $T$, then $T$ gives the red point $\pi(S(T))$ of its span and the blue point $\pi(G)$ of its inner gap $G$ (the gap that does not contain $\varphi$). If $\varphi$ lies in an arc of $T$, its red corner is always $(0,0)$, which dominates no blue point, so it can be counted for free and is not stored, while $\theta$ can lie in either gap, so $T$ gives the blue points $\pi(G)$ and $\pi(G')$ of its two gaps. In total, $\Pi$ has exactly $2n$ points (Fig.~\ref{fig:ctpoints}).

\begin{lemma}\label{lem:quadrants}
For every $\theta$, the corners of the boxes of $\T_\theta$ are the points of $\Pi\cap Q_\theta$, together with a red point at $(0,0)$ for each $T$ such that $\varphi$ lies in an arc of $T$, and a blue point at the apex $(\theta,1-\theta)$ for each $T$ such that $\theta$ lies in an arc of $T$.
\end{lemma}

\begin{proof}
By~\eqref{eq:pi}, a red point $\pi(S(T))$ lies in $Q_\theta$ if and only if $\theta\in S(T)$, that is, by Lemma~\ref{lem:span}, if and only if $T\in\T_\theta$, and it is then the red corner of $T$. A blue point $\pi(G)$ lies in $Q_\theta$ if and only if $\theta\in G$. Then $T\in\T_\theta$, since $G$ is the inner gap or $\varphi$ lies in an arc of $T$, and $\pi(G)$ is the blue corner of $T$. Conversely, a red corner other than $(0,0)$ and a blue corner other than the apex are of these two kinds, by Observation~\ref{obs:ends}.
\end{proof}

We call the corners at $(0,0)$ and at the apex, which are not points of $\Pi$, the \emph{free} corners. Each of them is compatible with every other corner: a red corner at $(0,0)$ dominates no blue corner, because blue corners have positive coordinates, and no red corner dominates the apex, since they lie in $Q_\theta$. So a set of corners is compatible if and only if its points in $\Pi$ are. Let $F_\theta$ be the total weight of the free corners, that is, the weight of the c-trapezoids with $\varphi$ in an arc plus the weight of the c-trapezoids with $\theta$ in an arc; then we have
\begin{equation}\label{eq:free}
\mwcs(\text{corners of the boxes of } \T_\theta) \;=\; \mwcs(\Pi\cap Q_\theta)+F_\theta .
\end{equation}

\begin{theorem}\label{thm:ct}
The maximum weight of a clique of $n$ circle trapezoids equals
\[
\max_\theta\ \bigl(\mwcs(\Pi\cap Q_\theta)+F_\theta-W_\theta\bigr),
\]
over one $\theta$ in each of the $4n+1$ intervals into which the endpoints cut $(0,1)$. Consequently, if Problem~\ref{prob:rb} on $m$ points can be preprocessed in time $P(m)$ so that each query takes time $Q(m)$, then a maximum weight clique of $n$ circle trapezoids can be found in time $O(P(2n)+n\,Q(2n)+n\log n)$.
\end{theorem}

\begin{proof}
By Lemma~\ref{lem:span}, the set $\T_\theta$ does not change while $\theta$ stays in one of the $4n+1$ intervals, and by Lemma~\ref{lem:chord}, every clique is contained in $\T_\theta$ for some $\theta$ that is not an endpoint. So the maximum weight of a clique is the maximum, over the chosen values of $\theta$, of the maximum weight of a clique of $\T_\theta$, which is $\mwcs(\Pi\cap Q_\theta)+F_\theta-W_\theta$ by~\eqref{eq:theta} and~\eqref{eq:free}.

For the running time, we sort the $4n$ endpoints in $O(n\log n)$ time, which gives the intervals and one $\theta$ in each, and build $\Pi$ in $O(n)$ more time. We compute all values $W_\theta$ and $F_\theta$ in one sweep over the intervals. In the first interval, $\theta$ lies before every endpoint, so it lies in an arc of a c-trapezoid $T$ exactly when $\varphi$ does. Hence both values count only the c-trapezoids $T$ with $\varphi$ in an arc: $W_\theta$ once, since $S(T)=(0,1)$, and $F_\theta$ twice. Moving to the next interval, $\theta$ passes one endpoint of a c-trapezoid $T$, so whether $\theta$ lies in the span or in an arc of any other c-trapezoid does not change. As for $T$, $\theta$ enters or leaves an arc of $T$, so $F_\theta$ gains or loses $w(T)$, and if the endpoint is $m(T)$ or $M(T)$, then $\theta$ enters or leaves $S(T)$, so $W_\theta$ gains or loses $w(T)$ as well. Each step takes constant time, so all values take $O(n)$ time. We then build the structure for Problem~\ref{prob:rb} on $\Pi$ in time $P(2n)$ and ask the $4n+1$ queries in time $O(n\,Q(2n))$, which gives the maximum weight of a clique.

To find a clique of this weight, let $\theta^*$ attain the maximum. A maximum weight clique of $\T_{\theta^*}$ is then a maximum weight clique of all the c-trapezoids. By Lemma~\ref{lem:lines}, $\T_{\theta^*}$ is a trapezoid graph, given by the intervals $[x_1,x_2]$ on $L_1$ and $[y_1,y_2]$ on $L_2$ of its c-trapezoids, which Observation~\ref{obs:ends} gives in $O(n)$ time. So the clique can be recovered in $O(n\log n)$ time using the algorithm of \cite{FMW97}.
\end{proof}

\section{The reduction for polygon-circle graphs}\label{sec:pc}

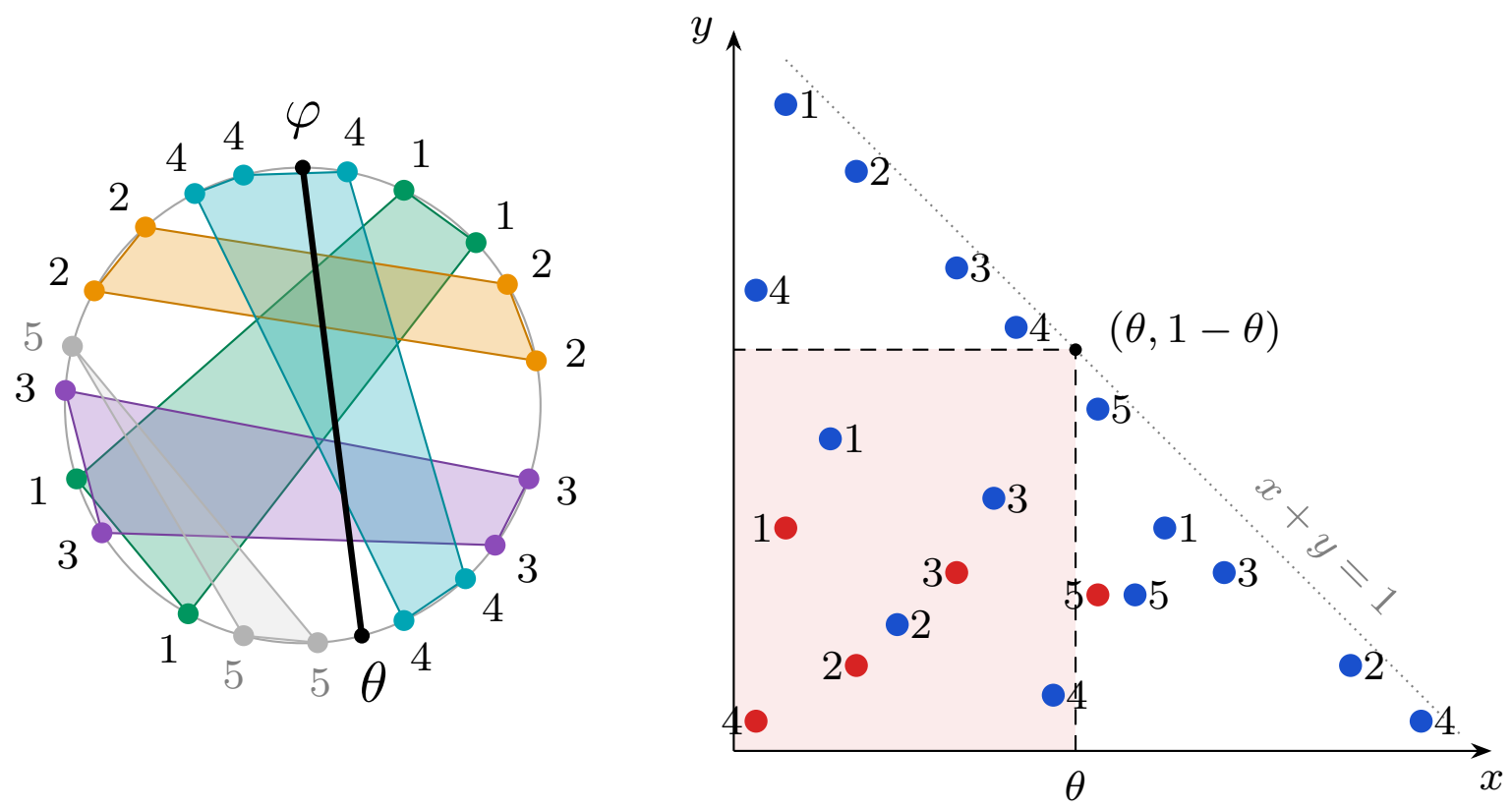
\begin{figure}[htb]
\centering
\input{figs/pc_fig_summary.tex}
\caption{(a)~Five inscribed polygons and a chord $\varphi\theta$, which intersects four of them. The numbers mark the vertices. (b)~The point set $\Pi$, one point per vertex, and the quadrant $Q_\theta$. The points in the quadrant are exactly the corners of the boxes of the four polygons intersected by the chord.}
\label{fig:pc}
\end{figure}

An \emph{inscribed polygon} is the convex hull $P$ of $k\ge2$ points of $C$, its \emph{vertices}; for $k=2$ it is a chord. Removing the vertices from $C$ leaves $k$ open arcs, the gaps of $P$. We are given $n$ polygons with $N$ vertices in total, each with a weight, and assume that the vertices are pairwise distinct and that $\varphi$ is not a vertex. The reduction of Section~\ref{sec:ct} applies with vertices in place of arcs, and it becomes simpler: $\varphi$ and $\theta$ never lie in a vertex, so every corner of a box will be a point of the set $\Pi$ defined below. We go through the steps and point out the changes (Fig.~\ref{fig:pc}).

\paragraph{Disjointness, spans and the chord.}
Lemma~\ref{lem:gap} holds with ``all vertices of $P'$'' in place of ``both arcs of $T'$''; there are now $k$ regions $R_G$, one for each gap $G$, bounded by $G$ and the side of $P$ that joins its ends, and the proof is unchanged. Let $v_1<\dots<v_k$ be the positions of the vertices of $P$. The span is $S(P)=[m(P),M(P)]=[v_1,v_k]$. The gap between $v_k$ and $v_1$ contains $\varphi$, and the other $k-1$ gaps, $(v_i,v_{i+1})$ for $i<k$, lie inside the span; we call them the \emph{inner gaps}. Lemmas~\ref{lem:span} and~\ref{lem:chord} hold verbatim, so to find a maximum weight clique it suffices to try one $\theta$ in each of the $N+1$ intervals into which the vertices cut $(0,1)$.

\medskip
Let $\PP_\theta$ be the set of polygons intersected by the chord $\varphi\theta$. For $P\in\PP_\theta$, Observation~\ref{obs:ends} becomes: $x_1=v_1$ and $y_1=1-v_k$, and $x_2=p$ and $y_2=1-q$ for the inner gap $(p,q)$ containing $\theta$. Lemma~\ref{lem:lines} holds with the same proof, and so does~\eqref{eq:theta} with $\PP_\theta$ in place of $\T_\theta$.

\paragraph{The point set.}
The red corner of $P$ is $\pi(S(P))=(v_1,1-v_k)$, and its blue corner is $\pi(G)$ for the inner gap $G$ containing $\theta$. We let $\Pi$ consist, for every polygon, of the red point $\pi(S(P))$ and the blue points $\pi((v_i,v_{i+1}))=(v_i,1-v_{i+1})$ for $i=1,\dots,k-1$, with the weight of the polygon. So every polygon gives one point per vertex, and $\Pi$ has exactly $N$ points. The proof of Lemma~\ref{lem:quadrants} now shows that the corners of the boxes of $\PP_\theta$ are exactly the points of $\Pi\cap Q_\theta$. Since the inner gaps of a polygon are disjoint, a quadrant contains at most one blue point of each polygon, and as $\theta$ moves, the blue corner of a polygon moves from one of its blue points to the next.

\begin{theorem}\label{thm:pc}
The maximum weight of a clique of $n$ inscribed polygons with $N$ vertices in total equals
\[
\max_\theta\ \bigl(\mwcs(\Pi\cap Q_\theta)-W_\theta\bigr),
\]
over one $\theta$ in each of the $N+1$ intervals into which the vertices cut $(0,1)$. Consequently, if Problem~\ref{prob:rb} on $m$ points can be preprocessed in time $P(m)$ so that each query takes time $Q(m)$, then a maximum weight clique of the polygons can be found in time $O(P(N)+N\,Q(N)+N\log N)$.
\end{theorem}

\begin{proof}
By Lemmas~\ref{lem:span} and~\ref{lem:chord}, which hold verbatim, the set $\PP_\theta$ does not change while $\theta$ stays in one of the $N+1$ intervals, and every clique is contained in $\PP_\theta$ for some $\theta$ that is not a vertex. For each such $\theta$, the corners of the boxes of $\PP_\theta$ are exactly the points of $\Pi\cap Q_\theta$, so by~\eqref{eq:theta} the maximum weight of a clique of $\PP_\theta$ is $\mwcs(\Pi\cap Q_\theta)-W_\theta$. This proves the formula.

For the running time, we sort the $N$ vertices in $O(N\log N)$ time, which gives the intervals and one $\theta$ in each, and build $\Pi$ in $O(N)$ more time. In the first interval, $\theta$ lies before every vertex, so it lies in no span and $W_\theta=0$. Moving to the next interval, $\theta$ passes one vertex of a polygon $P$, and $W_\theta$ gains or loses $w(P)$ if this vertex is $v_1$ or $v_k$, where $\theta$ enters or leaves $S(P)$, and does not change otherwise. So all values $W_\theta$ take $O(N)$ time. We then build the structure for Problem~\ref{prob:rb} on $\Pi$ in time $P(N)$ and ask the $N+1$ queries in time $O(N\,Q(N))$, which gives the maximum weight of a clique.

To find a clique of this weight, let $\theta^*$ attain the maximum. As in the proof of Theorem~\ref{thm:ct}, a maximum weight clique of $\PP_{\theta^*}$ is a maximum weight clique of all the polygons, and by Lemma~\ref{lem:lines}, $\PP_{\theta^*}$ is a trapezoid graph, given by the intervals $[x_1,x_2]$ on $L_1$ and $[y_1,y_2]$ on $L_2$ of its polygons. For a polygon with $k$ vertices, these are $x_1=v_1$, $y_1=1-v_k$, $x_2=p$ and $y_2=1-q$ for the inner gap $(p,q)$ containing $\theta^*$, which we find in $O(k)$ time.\footnote{Or even in $O(\log k)$ time by binary search, if the vertices of the polygon are sorted by position. This does not change the total bound.} So all intervals take $O(N)$ time. Then the clique can be recovered in $O(n\log n)$ time using the algorithm of \cite{FMW97}.
\end{proof}

\begin{remark}
The two reductions combine verbatim for convex hulls of finitely many disjoint points and arcs of $C$, with one point of $\Pi$ per point or arc and free corners for the arcs.
As graph classes, circle trapezoid graphs are polygon-circle graphs, but we do not obtain Theorem~\ref{thm:main} from Theorem~\ref{thm:poly}. The obvious way to replace the arcs of a c-trapezoid model by vertices while preserving the graph puts a vertex next to every endpoint of another c-trapezoid inside each arc, which can produce $\Theta(n^2)$ vertices, and we do not know whether every circle trapezoid graph has a polygon model of linear size.
\end{remark}

\input{bichromaticqueries}

\section{Assembling the results}\label{sec:assembl}

\mainthm*

\begin{proof}
Theorem~\ref{thm:ct} with Theorem~\ref{thm:ds}, that is, with $P(m)=O(m\log^2m)$ and $Q(m)=O(\log m)$, finds a maximum weight clique of $n$ circle trapezoids in $O(n\log^2 n)$ time.
\end{proof}

\circlecor*

\begin{proof}
Circle graphs are circle trapezoid graphs as well: replace each chord $ab$ by the c-trapezoid with arcs $[a,a+\varepsilon]$
and $[b,b+\varepsilon]$, for $\varepsilon$ smaller than the distance between any two endpoints, and apply Theorem~\ref{thm:main}.\footnote{Since chords are much simpler objects than c-trapezoids, a similar approach can also be carried out directly on the chords, which gives a simpler algorithm with the same running time.}
\end{proof}

\arccor*

\begin{proof}
Circular-arc graphs are circle trapezoid graphs with models of the same size: replace each arc $[s,e]$ by the c-trapezoid with arcs $[s,e-2\varepsilon]$ and $[e-\varepsilon,e]$, for $\varepsilon$ smaller than half the distance between any two endpoints, and apply Theorem~\ref{thm:main}. Alternatively, the reduction of \cite[Section~8]{MPS27} for the weighted problem, with Theorem~\ref{thm:ds} answering its quadrant queries, takes $O(n\log^2 n)$ time as well.
\end{proof}

\polythm*

\begin{proof}
Theorem~\ref{thm:pc} with Theorem~\ref{thm:ds} finds a maximum weight clique of $n$ polygons with $N$ vertices in $O(N\log^2 N)$ time.
\end{proof}
\bibliographystyle{alphaurl}
\bibliography{references}

\section*{AI Disclosure.}
Generative AI tools (Claude, ChatGPT) were used to formalize the technical details and to generate the figures. The authors assume responsibility for all content.


\end{document}

%% file: figs/fig_overview.tex
\begin{tikzpicture}
\begin{scope}
\draw[gray!70] (0,0) circle (1.25);
\fill[tA,opacity=.28] (75.6:1.25) arc[start angle=75.6, end angle=36, radius=1.25] -- (-126:1.25) arc[start angle=-126, end angle=-151.2, radius=1.25] -- cycle;
\draw[tA!85!black,thin] (75.6:1.25) arc[start angle=75.6, end angle=36, radius=1.25] -- (-126:1.25) arc[start angle=-126, end angle=-151.2, radius=1.25] -- cycle;
\fill[tB,opacity=.28] (50.4:1.25) arc[start angle=50.4, end angle=21.6, radius=1.25] -- (-208.8:1.25) arc[start angle=-208.8, end angle=-234, radius=1.25] -- cycle;
\draw[tB!85!black,thin] (50.4:1.25) arc[start angle=50.4, end angle=21.6, radius=1.25] -- (-208.8:1.25) arc[start angle=-208.8, end angle=-234, radius=1.25] -- cycle;
\fill[tC,opacity=.28] (3.6:1.25) arc[start angle=3.6, end angle=-18, radius=1.25] -- (-162:1.25) arc[start angle=-162, end angle=-183.6, radius=1.25] -- cycle;
\draw[tC!85!black,thin] (3.6:1.25) arc[start angle=3.6, end angle=-18, radius=1.25] -- (-162:1.25) arc[start angle=-162, end angle=-183.6, radius=1.25] -- cycle;
\fill[tD,opacity=.28] (-248.4:1.25) arc[start angle=-248.4, end angle=-280.8, radius=1.25] -- (-39.6:1.25) arc[start angle=-39.6, end angle=-54, radius=1.25] -- cycle;
\draw[tD!85!black,thin] (-248.4:1.25) arc[start angle=-248.4, end angle=-280.8, radius=1.25] -- (-39.6:1.25) arc[start angle=-39.6, end angle=-54, radius=1.25] -- cycle;
\fill[gray!40,opacity=.25] (-79.2:1.25) arc[start angle=-79.2, end angle=-97.2, radius=1.25] -- (-190.8:1.25) arc[start angle=-190.8, end angle=-201.6, radius=1.25] -- cycle;
\draw[gray!60] (-79.2:1.25) arc[start angle=-79.2, end angle=-97.2, radius=1.25] -- (-190.8:1.25) arc[start angle=-190.8, end angle=-201.6, radius=1.25] -- cycle;
\draw[tA,line width=1.8pt] (75.6:1.25) arc[start angle=75.6, end angle=36, radius=1.25];
\draw[tA,line width=1.8pt] (-126:1.25) arc[start angle=-126, end angle=-151.2, radius=1.25];
\draw[tB,line width=1.8pt] (50.4:1.34) arc[start angle=50.4, end angle=21.6, radius=1.34];
\draw[tB,line width=1.8pt] (-208.8:1.25) arc[start angle=-208.8, end angle=-234, radius=1.25];
\draw[tC,line width=1.8pt] (3.6:1.25) arc[start angle=3.6, end angle=-18, radius=1.25];
\draw[tC,line width=1.8pt] (-162:1.25) arc[start angle=-162, end angle=-183.6, radius=1.25];
\draw[tD,line width=1.8pt] (-248.4:1.25) arc[start angle=-248.4, end angle=-280.8, radius=1.25];
\draw[tD,line width=1.8pt] (-39.6:1.25) arc[start angle=-39.6, end angle=-54, radius=1.25];
\draw[gray!60,line width=1.8pt] (-79.2:1.25) arc[start angle=-79.2, end angle=-97.2, radius=1.25];
\draw[gray!60,line width=1.8pt] (-190.8:1.25) arc[start angle=-190.8, end angle=-201.6, radius=1.25];
\draw[very thick] (90:1.25) -- (-68.4:1.25);
\fill (90:1.25) circle (1.3pt); \fill (-68.4:1.25) circle (1.3pt);
\node at (90:1.52) {$\varphi$};
\node at (-68.4:1.52) {$\theta$};
\node at (0,-2) {\small (a)};
\end{scope}
\draw[-{Stealth[length=6pt]},thick,gray!70!black] (1.51,0) -- (2.21,0);
\begin{scope}[xshift=2.527cm,yshift=-0.8cm]
\fill[tA,opacity=.28] (0.227,1.6) -- (0.852,1.6) -- (1.786,0) -- (1.473,0) -- cycle;
\draw[tA!85!black,thin] (0.227,1.6) -- (0.852,1.6) -- (1.786,0) -- (1.473,0) -- cycle;
\fill[tB,opacity=.28] (0.625,1.6) -- (1.08,1.6) -- (0.759,0) -- (0.446,0) -- cycle;
\draw[tB!85!black,thin] (0.625,1.6) -- (1.08,1.6) -- (0.759,0) -- (0.446,0) -- cycle;
\fill[tC,opacity=.28] (1.364,1.6) -- (1.705,1.6) -- (1.339,0) -- (1.071,0) -- cycle;
\draw[tC!85!black,thin] (1.364,1.6) -- (1.705,1.6) -- (1.339,0) -- (1.071,0) -- cycle;
\fill[tD,opacity=.28] (0,1.6) -- (2.273,1.6) -- (0.268,0) -- (0,0) -- cycle;
\draw[tD!85!black,thin] (0,1.6) -- (2.273,1.6) -- (0.268,0) -- (0,0) -- cycle;
\draw[thin] (0,1.6) -- (2.5,1.6);
\draw[thin] (0,0) -- (2.5,0);
\draw[tA,line width=1.8pt] (0.227,1.6) -- (0.852,1.6);
\draw[tA,line width=1.8pt] (1.473,0) -- (1.786,0);
\draw[tB,line width=1.8pt] (0.625,1.69) -- (1.08,1.69);
\draw[tB,line width=1.8pt] (0.446,0) -- (0.759,0);
\draw[tC,line width=1.8pt] (1.364,1.6) -- (1.705,1.6);
\draw[tC,line width=1.8pt] (1.071,0) -- (1.339,0);
\draw[tD,line width=1.8pt] (0,0) -- (0.268,0);
\draw[tD,line width=1.8pt] (0,1.6) -- (0.17,1.6);
\draw[tD,line width=1.8pt] (2.045,1.6) -- (2.273,1.6);
\fill (0,1.6) circle (1.1pt); \fill (0,0) circle (1.1pt);
\fill (2.5,1.6) circle (1.1pt); \fill (2.5,0) circle (1.1pt);
\node[above] at (0,1.68) {$\varphi$}; \node[below] at (0,-0.05) {$\varphi$};
\node[above] at (2.5,1.68) {$\theta$}; \node[below] at (2.5,-0.05) {$\theta$};
\node at (1.25,-1.2) {\small (b)};
\end{scope}
\draw[-{Stealth[length=6pt]},thick,gray!70!black] (5.345,0) -- (6.045,0);
\begin{scope}[xshift=6.328cm,yshift=-1.4cm]
\fill[redpt!11] (0,0) -- (0,1.825) -- (0.325,1.825) -- (0.325,0.675) -- (0.8,0.675) -- (0.8,0.15) -- (2.2,0.15) -- (2.2,0) -- cycle;
\fill[bluept!10] (0,2.8) -- (0,1.825) -- (0.325,1.825) -- (0.325,0.675) -- (0.8,0.675) -- (0.8,0.15) -- (2.2,0.15) -- (2.2,2.8) -- cycle;
\fill[tA,opacity=.3] (0.2,1.65) rectangle (0.75,2); \draw[tA!85!black] (0.2,1.65) rectangle (0.75,2);
\fill[tB,opacity=.3] (0.55,0.5) rectangle (0.95,0.85); \draw[tB!85!black] (0.55,0.5) rectangle (0.95,0.85);
\fill[tC,opacity=.3] (1.2,1.2) rectangle (1.5,1.5); \draw[tC!85!black] (1.2,1.2) rectangle (1.5,1.5);
\fill[tD,opacity=.3] (0,0) rectangle (2,0.3); \draw[tD!85!black] (0,0) rectangle (2,0.3);
\draw[very thick] (0,1.825) -- (0.325,1.825) -- (0.325,0.675) -- (0.8,0.675) -- (0.8,0.15) -- (2.2,0.15);
\draw (0,0) rectangle (2.2,2.8);
\node[rps] at (0.2,1.65) {}; \node[bps] at (0.75,2) {};
\node[rps] at (0.55,0.5) {}; \node[bps] at (0.95,0.85) {};
\node[rps] at (1.2,1.2) {}; \node[bps] at (1.5,1.5) {};
\node[rps] at (0,0) {}; \node[bps] at (2,0.3) {};
\node at (1.1,-0.6) {\small (c)};
\end{scope}
\draw[-{Stealth[length=6pt]},thick,gray!70!black] (8.785,0) -- (9.485,0);
\begin{scope}[xshift=9.793cm,yshift=-1.476cm]
\fill[redpt!9] (0,0) rectangle (1.804,2.296);
\draw[densely dashed] (1.804,0) -- (1.804,2.296) -- (0,2.296);
\draw[densely dotted,gray] (1.148,2.952) -- (2.542,1.558);
\draw[-{Stealth[length=3pt]}] (0,0) -- (2.692,0);
\draw[-{Stealth[length=3pt]}] (0,0) -- (0,3.102);
\node[rps] at (0.164,1.353) {};
\node[bps] at (0.615,1.64) {};
\node[rps] at (0.451,0.41) {};
\node[bps] at (0.779,0.697) {};
\node[rps] at (0.984,0.984) {};
\node[bps] at (1.23,1.23) {};
\node[bps] at (0.123,2.624) {};
\node[bps] at (1.64,0.246) {};
\node[rps] at (1.927,0.779) {};
\node[bps] at (2.132,0.902) {};
\node[rpo,minimum size=4pt] at (0,0) {};
\fill (1.804,2.296) circle (1pt);
\node at (1.271,-0.524) {\small (d)};
\end{scope}
\end{tikzpicture}

%% file: figs/ct_fig_model.tex
\begin{tikzpicture}
\begin{scope}
\draw[gray!70] (0,0) circle (1.75);
\fill[tA,opacity=.28] (75.6:1.75) arc[start angle=75.6, end angle=36, radius=1.75] -- (-126:1.75) arc[start angle=-126, end angle=-151.2, radius=1.75] -- cycle;
\draw[tA!85!black,thin] (75.6:1.75) arc[start angle=75.6, end angle=36, radius=1.75] -- (-126:1.75) arc[start angle=-126, end angle=-151.2, radius=1.75] -- cycle;
\fill[tB,opacity=.28] (50.4:1.75) arc[start angle=50.4, end angle=21.6, radius=1.75] -- (-208.8:1.75) arc[start angle=-208.8, end angle=-234, radius=1.75] -- cycle;
\draw[tB!85!black,thin] (50.4:1.75) arc[start angle=50.4, end angle=21.6, radius=1.75] -- (-208.8:1.75) arc[start angle=-208.8, end angle=-234, radius=1.75] -- cycle;
\fill[tC,opacity=.28] (3.6:1.75) arc[start angle=3.6, end angle=-18, radius=1.75] -- (-162:1.75) arc[start angle=-162, end angle=-183.6, radius=1.75] -- cycle;
\draw[tC!85!black,thin] (3.6:1.75) arc[start angle=3.6, end angle=-18, radius=1.75] -- (-162:1.75) arc[start angle=-162, end angle=-183.6, radius=1.75] -- cycle;
\fill[tD,opacity=.28] (-248.4:1.75) arc[start angle=-248.4, end angle=-280.8, radius=1.75] -- (-39.6:1.75) arc[start angle=-39.6, end angle=-54, radius=1.75] -- cycle;
\draw[tD!85!black,thin] (-248.4:1.75) arc[start angle=-248.4, end angle=-280.8, radius=1.75] -- (-39.6:1.75) arc[start angle=-39.6, end angle=-54, radius=1.75] -- cycle;
\fill[tE,opacity=.28] (-79.2:1.75) arc[start angle=-79.2, end angle=-97.2, radius=1.75] -- (-190.8:1.75) arc[start angle=-190.8, end angle=-201.6, radius=1.75] -- cycle;
\draw[tE!85!black,thin] (-79.2:1.75) arc[start angle=-79.2, end angle=-97.2, radius=1.75] -- (-190.8:1.75) arc[start angle=-190.8, end angle=-201.6, radius=1.75] -- cycle;
\draw[tA,line width=2.4pt] (75.6:1.75) arc[start angle=75.6, end angle=36, radius=1.75];
\node[lab,text=black] at (55.8:2.03) {1};
\draw[tA,line width=2.4pt] (-126:1.75) arc[start angle=-126, end angle=-151.2, radius=1.75];
\node[lab,text=black] at (-138.6:2.03) {1};
\draw[tB,line width=2.4pt] (50.4:1.84) arc[start angle=50.4, end angle=21.6, radius=1.84];
\node[lab,text=black] at (36:2.03) {2};
\draw[tB,line width=2.4pt] (-208.8:1.75) arc[start angle=-208.8, end angle=-234, radius=1.75];
\node[lab,text=black] at (-221.4:2.03) {2};
\draw[tC,line width=2.4pt] (3.6:1.75) arc[start angle=3.6, end angle=-18, radius=1.75];
\node[lab,text=black] at (-7.2:2.03) {3};
\draw[tC,line width=2.4pt] (-162:1.75) arc[start angle=-162, end angle=-183.6, radius=1.75];
\node[lab,text=black] at (-172.8:2.03) {3};
\draw[tD,line width=2.4pt] (-248.4:1.75) arc[start angle=-248.4, end angle=-280.8, radius=1.75];
\node[lab,text=black] at (-254.52:2.03) {4};
\draw[tD,line width=2.4pt] (-39.6:1.75) arc[start angle=-39.6, end angle=-54, radius=1.75];
\node[lab,text=black] at (-46.8:2.03) {4};
\draw[tE,line width=2.4pt] (-79.2:1.75) arc[start angle=-79.2, end angle=-97.2, radius=1.75];
\node[lab,text=black] at (-88.2:2.03) {5};
\draw[tE,line width=2.4pt] (-190.8:1.75) arc[start angle=-190.8, end angle=-201.6, radius=1.75];
\node[lab,text=black] at (-196.2:2.03) {5};
\node at (0,-2.75) {\small (a)};
\end{scope}
\begin{scope}[xshift=6.2cm]
\fill[gray!20] (21.6:1.75) arc[start angle=21.6, end angle=-208.8, radius=1.75] -- cycle;
\fill[gray!20] (-234:1.75) arc[start angle=-234, end angle=-309.6, radius=1.75] -- cycle;
\draw[gray!70] (0,0) circle (1.75);
\fill[tB,opacity=.28] (50.4:1.75) arc[start angle=50.4, end angle=21.6, radius=1.75] -- (-208.8:1.75) arc[start angle=-208.8, end angle=-234, radius=1.75] -- cycle;
\draw[tB!85!black,thin] (50.4:1.75) arc[start angle=50.4, end angle=21.6, radius=1.75] -- (-208.8:1.75) arc[start angle=-208.8, end angle=-234, radius=1.75] -- cycle;
\fill[tA,opacity=.28] (75.6:1.75) arc[start angle=75.6, end angle=36, radius=1.75] -- (-126:1.75) arc[start angle=-126, end angle=-151.2, radius=1.75] -- cycle;
\draw[tA!85!black,thin] (75.6:1.75) arc[start angle=75.6, end angle=36, radius=1.75] -- (-126:1.75) arc[start angle=-126, end angle=-151.2, radius=1.75] -- cycle;
\fill[tC,opacity=.28] (3.6:1.75) arc[start angle=3.6, end angle=-18, radius=1.75] -- (-162:1.75) arc[start angle=-162, end angle=-183.6, radius=1.75] -- cycle;
\draw[tC!85!black,thin] (3.6:1.75) arc[start angle=3.6, end angle=-18, radius=1.75] -- (-162:1.75) arc[start angle=-162, end angle=-183.6, radius=1.75] -- cycle;
\draw[tB,line width=2.4pt] (50.4:1.84) arc[start angle=50.4, end angle=21.6, radius=1.84];
\node[lab,text=black] at (36:2.03) {2};
\draw[tB,line width=2.4pt] (-208.8:1.75) arc[start angle=-208.8, end angle=-234, radius=1.75];
\node[lab,text=black] at (-221.4:2.03) {2};
\draw[tA,line width=2.4pt] (75.6:1.75) arc[start angle=75.6, end angle=36, radius=1.75];
\node[lab,text=black] at (55.8:2.03) {1};
\draw[tA,line width=2.4pt] (-126:1.75) arc[start angle=-126, end angle=-151.2, radius=1.75];
\node[lab,text=black] at (-138.6:2.03) {1};
\draw[tC,line width=2.4pt] (3.6:1.75) arc[start angle=3.6, end angle=-18, radius=1.75];
\node[lab,text=black] at (-7.2:2.03) {3};
\draw[tC,line width=2.4pt] (-162:1.75) arc[start angle=-162, end angle=-183.6, radius=1.75];
\node[lab,text=black] at (-172.8:2.03) {3};
\draw[densely dotted,thick] (19.44:2.3) arc[start angle=19.44, end angle=-206.64, radius=2.3];
\node[font=\small] at (-43.2:2.57) {$G$};
\draw[densely dotted,thick] (-236.16:2.3) arc[start angle=-236.16, end angle=-307.44, radius=2.3];
\node[font=\small] at (84.6:2.57) {$G'$};
\node at (0,-2.75) {\small (b)};
\end{scope}
\end{tikzpicture}

%% file: figs/ct_fig_chord.tex
\begin{tikzpicture}
\begin{scope}
\draw[gray!70] (0,0) circle (1.75);
\fill[tA,opacity=.28] (75.6:1.75) arc[start angle=75.6, end angle=36, radius=1.75] -- (-126:1.75) arc[start angle=-126, end angle=-151.2, radius=1.75] -- cycle;
\draw[tA!85!black,thin] (75.6:1.75) arc[start angle=75.6, end angle=36, radius=1.75] -- (-126:1.75) arc[start angle=-126, end angle=-151.2, radius=1.75] -- cycle;
\fill[tB,opacity=.28] (50.4:1.75) arc[start angle=50.4, end angle=21.6, radius=1.75] -- (-208.8:1.75) arc[start angle=-208.8, end angle=-234, radius=1.75] -- cycle;
\draw[tB!85!black,thin] (50.4:1.75) arc[start angle=50.4, end angle=21.6, radius=1.75] -- (-208.8:1.75) arc[start angle=-208.8, end angle=-234, radius=1.75] -- cycle;
\fill[tC,opacity=.28] (3.6:1.75) arc[start angle=3.6, end angle=-18, radius=1.75] -- (-162:1.75) arc[start angle=-162, end angle=-183.6, radius=1.75] -- cycle;
\draw[tC!85!black,thin] (3.6:1.75) arc[start angle=3.6, end angle=-18, radius=1.75] -- (-162:1.75) arc[start angle=-162, end angle=-183.6, radius=1.75] -- cycle;
\fill[tD,opacity=.28] (-248.4:1.75) arc[start angle=-248.4, end angle=-280.8, radius=1.75] -- (-39.6:1.75) arc[start angle=-39.6, end angle=-54, radius=1.75] -- cycle;
\draw[tD!85!black,thin] (-248.4:1.75) arc[start angle=-248.4, end angle=-280.8, radius=1.75] -- (-39.6:1.75) arc[start angle=-39.6, end angle=-54, radius=1.75] -- cycle;
\fill[gray!40,opacity=.25] (-79.2:1.75) arc[start angle=-79.2, end angle=-97.2, radius=1.75] -- (-190.8:1.75) arc[start angle=-190.8, end angle=-201.6, radius=1.75] -- cycle;
\draw[gray!60] (-79.2:1.75) arc[start angle=-79.2, end angle=-97.2, radius=1.75] -- (-190.8:1.75) arc[start angle=-190.8, end angle=-201.6, radius=1.75] -- cycle;
\draw[tA,line width=2.4pt] (75.6:1.75) arc[start angle=75.6, end angle=36, radius=1.75];
\node[lab,text=black] at (55.8:2.03) {1};
\draw[tA,line width=2.4pt] (-126:1.75) arc[start angle=-126, end angle=-151.2, radius=1.75];
\node[lab,text=black] at (-138.6:2.03) {1};
\draw[tB,line width=2.4pt] (50.4:1.84) arc[start angle=50.4, end angle=21.6, radius=1.84];
\node[lab,text=black] at (36:2.03) {2};
\draw[tB,line width=2.4pt] (-208.8:1.75) arc[start angle=-208.8, end angle=-234, radius=1.75];
\node[lab,text=black] at (-221.4:2.03) {2};
\draw[tC,line width=2.4pt] (3.6:1.75) arc[start angle=3.6, end angle=-18, radius=1.75];
\node[lab,text=black] at (-7.2:2.03) {3};
\draw[tC,line width=2.4pt] (-162:1.75) arc[start angle=-162, end angle=-183.6, radius=1.75];
\node[lab,text=black] at (-172.8:2.03) {3};
\draw[tD,line width=2.4pt] (-248.4:1.75) arc[start angle=-248.4, end angle=-280.8, radius=1.75];
\node[lab,text=black] at (-254.52:2.03) {4};
\draw[tD,line width=2.4pt] (-39.6:1.75) arc[start angle=-39.6, end angle=-54, radius=1.75];
\node[lab,text=black] at (-46.8:2.03) {4};
\draw[gray!60,line width=2.4pt] (-79.2:1.75) arc[start angle=-79.2, end angle=-97.2, radius=1.75];
\node[lab,text=gray] at (-88.2:2.03) {5};
\draw[gray!60,line width=2.4pt] (-190.8:1.75) arc[start angle=-190.8, end angle=-201.6, radius=1.75];
\node[lab,text=gray] at (-196.2:2.03) {5};
\draw[very thick] (90:1.75) -- (-68.4:1.75);
\fill (90:1.75) circle (1.5pt);
\fill (-68.4:1.75) circle (1.5pt);
\node at (90:2.07) {$\varphi$};
\node at (-68.4:2.07) {$\theta$};
\draw[-{Stealth[length=4pt]},gray!80!black] (77.4:2.37) arc[start angle=77.4, end angle=48.6, radius=2.37] node[right,lab,text=black] {$x$};
\draw[-{Stealth[length=4pt]},gray!80!black] (-257.4:2.37) arc[start angle=-257.4, end angle=-228.6, radius=2.37] node[left,lab,text=black] {$y$};
\node at (12.6:2.35) {$L_1$};
\node at (-158.4:2.35) {$L_2$};
\node at (0,-2.75) {\small (a)};
\end{scope}
\begin{scope}[xshift=3.55cm,yshift=-1.5cm]
\draw[thin] (0,0) -- (7,0);
\draw (0,-0.08) -- (0,0.08);
\draw (7,-0.08) -- (7,0.08);
\node[below] at (0,-0.05) {$\varphi$};
\node[below] at (7,-0.05) {$\varphi$};
\draw[tA,thin] (0.28,2.65) -- (4.69,2.65);
\draw[tA,line width=2.4pt] (0.28,2.65) -- (1.05,2.65);
\draw[tA,line width=2.4pt] (4.2,2.65) -- (4.69,2.65);
\node[lab,left] at (-0.12,2.65) {$T_1$};
\draw[tB,thin] (0.77,2.13) -- (6.3,2.13);
\draw[tB,line width=2.4pt] (0.77,2.13) -- (1.33,2.13);
\draw[tB,line width=2.4pt] (5.81,2.13) -- (6.3,2.13);
\node[lab,left] at (-0.12,2.13) {$T_2$};
\draw[tC,thin] (1.68,1.61) -- (5.32,1.61);
\draw[tC,line width=2.4pt] (1.68,1.61) -- (2.1,1.61);
\draw[tC,line width=2.4pt] (4.9,1.61) -- (5.32,1.61);
\node[lab,left] at (-0.12,1.61) {$T_3$};
\draw[tD,thin] (0,1.09) -- (7,1.09);
\draw[tD,line width=2.4pt] (6.58,1.09) -- (7,1.09);
\draw[tD,line width=2.4pt] (0,1.09) -- (0.21,1.09);
\draw[tD,line width=2.4pt] (2.52,1.09) -- (2.8,1.09);
\node[lab,left] at (-0.12,1.09) {$T_4$};
\draw[tE,thin] (3.29,0.57) -- (5.67,0.57);
\draw[tE,line width=2.4pt] (3.29,0.57) -- (3.64,0.57);
\draw[tE,line width=2.4pt] (5.46,0.57) -- (5.67,0.57);
\node[lab,left] at (-0.12,0.57) {$T_5$};
\draw[densely dashed] (3.08,-0.05) -- (3.08,2.95);
\node[below] at (3.08,-0.05) {$\theta$};
\draw[decorate,decoration={brace,amplitude=3pt,raise=4pt}] (0.28,2.65) -- (4.69,2.65) node[midway,above=7pt,lab] {$S(T_1)=[m(T_1),M(T_1)]$};
\node at (3.5,-1.25) {\small (b)};
\end{scope}
\end{tikzpicture}

%% file: figs/ct_fig_lines.tex
\begin{tikzpicture}
\begin{scope}
\fill[tA,opacity=.28] (0.582,2.5) -- (2.182,2.5) -- (4.571,0) -- (3.771,0) -- cycle;
\draw[tA!85!black,thin] (0.582,2.5) -- (2.182,2.5) -- (4.571,0) -- (3.771,0) -- cycle;
\fill[tB,opacity=.28] (1.6,2.5) -- (2.764,2.5) -- (1.943,0) -- (1.143,0) -- cycle;
\draw[tB!85!black,thin] (1.6,2.5) -- (2.764,2.5) -- (1.943,0) -- (1.143,0) -- cycle;
\fill[tC,opacity=.28] (3.491,2.5) -- (4.364,2.5) -- (3.429,0) -- (2.743,0) -- cycle;
\draw[tC!85!black,thin] (3.491,2.5) -- (4.364,2.5) -- (3.429,0) -- (2.743,0) -- cycle;
\fill[tD,opacity=.28] (0,2.5) -- (5.818,2.5) -- (0.686,0) -- (0,0) -- cycle;
\draw[tD!85!black,thin] (0,2.5) -- (5.818,2.5) -- (0.686,0) -- (0,0) -- cycle;
\draw[thin] (0,2.5) -- (6.4,2.5);
\draw[thin] (0,0) -- (6.4,0);
\draw[tA,line width=2.4pt] (0.582,2.5) -- (2.182,2.5);
\node[lab] at (1.382,2.77) {1};
\draw[tA,line width=2.4pt] (3.771,0) -- (4.571,0);
\node[lab] at (4.171,-0.27) {1};
\draw[tB,line width=2.4pt] (1.6,2.59) -- (2.764,2.59);
\node[lab] at (2.182,2.77) {2};
\draw[tB,line width=2.4pt] (1.143,0) -- (1.943,0);
\node[lab] at (1.543,-0.27) {2};
\draw[tC,line width=2.4pt] (3.491,2.5) -- (4.364,2.5);
\node[lab] at (3.927,2.77) {3};
\draw[tC,line width=2.4pt] (2.743,0) -- (3.429,0);
\node[lab] at (3.086,-0.27) {3};
\draw[tD,line width=2.4pt] (0,0) -- (0.686,0);
\node[lab] at (0.343,-0.27) {4};
\draw[tD,line width=2.4pt] (0,2.5) -- (0.436,2.5);
\node[lab] at (0.218,2.77) {4};
\draw[tD,line width=2.4pt] (5.236,2.5) -- (5.818,2.5);
\node[lab] at (5.527,2.77) {4};
\fill (0,2.5) circle (1.3pt);
\fill (0,0) circle (1.3pt);
\fill (6.4,2.5) circle (1.3pt);
\fill (6.4,0) circle (1.3pt);
\node[left] at (-0.03,2.5) {$\varphi$};
\node[left] at (-0.03,0) {$\varphi$};
\node[left] at (-0.45,2.5) {$L_1$};
\node[left] at (-0.45,0) {$L_2$};
\node[right] at (6.43,2.5) {$\theta$};
\node[right] at (6.43,0) {$\theta$};
\draw[-{Stealth[length=4pt]},gray!80!black] (2.2,3.15) -- (3.4,3.15) node[right,lab,text=black] {$x$};
\draw[-{Stealth[length=4pt]},gray!80!black] (2.2,-0.65) -- (3.4,-0.65) node[right,lab,text=black] {$y$};
\node at (3.2,-1.3) {\small (a)};
\end{scope}
\begin{scope}[xshift=8.4cm,yshift=-0.75cm]
\fill[redpt!11] (0,0) -- (0,2.847) -- (0.507,2.847) -- (0.507,1.053) -- (1.248,1.053) -- (1.248,0.234) -- (3.432,0.234) -- (3.432,0) -- cycle;
\fill[bluept!10] (0,4.368) -- (0,2.847) -- (0.507,2.847) -- (0.507,1.053) -- (1.248,1.053) -- (1.248,0.234) -- (3.432,0.234) -- (3.432,4.368) -- cycle;
\fill[tA,opacity=.3] (0.312,2.574) rectangle (1.17,3.12);
\draw[tA!85!black] (0.312,2.574) rectangle (1.17,3.12);
\fill[tB,opacity=.3] (0.858,0.78) rectangle (1.482,1.326);
\draw[tB!85!black] (0.858,0.78) rectangle (1.482,1.326);
\fill[tC,opacity=.3] (1.872,1.872) rectangle (2.34,2.34);
\draw[tC!85!black] (1.872,1.872) rectangle (2.34,2.34);
\fill[tD,opacity=.3] (0,0) rectangle (3.12,0.468);
\draw[tD!85!black] (0,0) rectangle (3.12,0.468);
\draw[very thick] (0,2.847) -- (0.507,2.847) -- (0.507,1.053) -- (1.248,1.053) -- (1.248,0.234) -- (3.432,0.234);
\draw[-{Stealth[length=4pt]}] (0,0) -- (3.782,0) node[below,lab] {$x$};
\draw[-{Stealth[length=4pt]}] (0,0) -- (0,4.718) node[left,lab] {$y$};
\draw[densely dashed,gray] (3.432,0) -- (3.432,4.368) -- (0,4.368);
\node[below,lab] at (3.432,0) {$\theta$};
\node[left,lab] at (0,4.368) {$1-\theta$};
\node[below left,lab] at (0,0) {$\varphi$};
\node[lab,above left] at (0.352,3.08) {1};
\node[rp] at (0.312,2.574) {};
\node[bp] at (1.17,3.12) {};
\node[lab,above left] at (0.898,1.286) {2};
\node[rp] at (0.858,0.78) {};
\node[bp] at (1.482,1.326) {};
\node[lab,above left] at (1.912,2.3) {3};
\node[rp] at (1.872,1.872) {};
\node[bp] at (2.34,2.34) {};
\node[lab,above] at (2.34,0.438) {4};
\node[rp] at (0,0) {};
\node[bp] at (3.12,0.468) {};
\node at (1.716,-0.55) {\small (b)};
\end{scope}
\end{tikzpicture}

%% file: figs/ct_fig_points.tex
\begin{tikzpicture}
\begin{scope}[xshift=0cm]
\fill[redpt!9] (0,0) rectangle (2.992,3.808);
\draw[densely dashed] (2.992,0) -- (2.992,3.808) -- (0,3.808);
\draw[densely dotted,gray] (1.904,4.896) -- (4.216,2.584) node[right,lab,text=gray] {$x+y=1$};
\draw[-{Stealth[length=4pt]}] (0,0) -- (4.416,0) node[below,lab] {$x$};
\draw[-{Stealth[length=4pt]}] (0,0) -- (0,5.096) node[left,lab] {$y$};
\node[below,lab] at (2.992,0) {$\theta$};
\node[rp] at (0.272,2.244) {};
\node[lab] at (0.102,2.244) {1};
\node[bp] at (1.02,2.72) {};
\node[lab] at (1.19,2.72) {1};
\node[rp] at (0.748,0.68) {};
\node[lab] at (0.578,0.68) {2};
\node[bp] at (1.292,1.156) {};
\node[lab] at (1.462,1.156) {2};
\node[rp] at (1.632,1.632) {};
\node[lab] at (1.462,1.632) {3};
\node[bp] at (2.04,2.04) {};
\node[lab] at (2.21,2.04) {3};
\node[bp] at (0.204,4.352) {};
\node[lab] at (0.374,4.352) {4};
\node[bp] at (2.72,0.408) {};
\node[lab] at (2.89,0.408) {4};
\node[rp] at (3.196,1.292) {};
\node[lab] at (3.026,1.292) {5};
\node[bp] at (3.536,1.496) {};
\node[lab] at (3.706,1.496) {5};
\node[rpo] at (0,0) {};
\fill (2.992,3.808) circle (1.2pt);
\node[lab,right] at (3.072,3.928) {$(\theta,1-\theta)$};
\node at (2.108,-0.75) {\small (a)};
\end{scope}
\begin{scope}[xshift=6.416cm]
\fill[redpt!9] (0,0) rectangle (3.332,3.468);
\draw[densely dashed] (3.332,0) -- (3.332,3.468) -- (0,3.468);
\draw[densely dotted,gray] (1.904,4.896) -- (4.216,2.584) node[right,lab,text=gray] {$x+y=1$};
\draw[-{Stealth[length=4pt]}] (0,0) -- (4.416,0) node[below,lab] {$x$};
\draw[-{Stealth[length=4pt]}] (0,0) -- (0,5.096) node[left,lab] {$y$};
\node[below,lab] at (3.332,0) {$\theta'$};
\node[rp] at (0.272,2.244) {};
\node[lab] at (0.102,2.244) {1};
\node[bp] at (1.02,2.72) {};
\node[lab] at (1.19,2.72) {1};
\node[rp] at (0.748,0.68) {};
\node[lab] at (0.578,0.68) {2};
\node[bp] at (1.292,1.156) {};
\node[lab] at (1.462,1.156) {2};
\node[rp] at (1.632,1.632) {};
\node[lab] at (1.462,1.632) {3};
\node[bp] at (2.04,2.04) {};
\node[lab] at (2.21,2.04) {3};
\node[bp] at (0.204,4.352) {};
\node[lab] at (0.374,4.352) {4};
\node[bp] at (2.72,0.408) {};
\node[lab] at (2.89,0.408) {4};
\node[rp] at (3.196,1.292) {};
\node[lab] at (3.026,1.292) {5};
\node[bp] at (3.536,1.496) {};
\node[lab] at (3.706,1.496) {5};
\node[rpo] at (0,0) {};
\node[bpo] at (3.332,3.468) {};
\node[lab,right] at (3.412,3.588) {$(\theta',1-\theta')$};
\node at (2.108,-0.75) {\small (b)};
\end{scope}
\end{tikzpicture}

%% file: figs/pc_fig_summary.tex
\begin{tikzpicture}
\begin{scope}
\draw[gray!70] (0,0) circle (1.6);
\fill[tA,opacity=.28] (64.8:1.6) -- (43.2:1.6) -- (-118.8:1.6) -- (-162:1.6) -- cycle;
\draw[tA!85!black,thin] (64.8:1.6) -- (43.2:1.6) -- (-118.8:1.6) -- (-162:1.6) -- cycle;
\fill[tB,opacity=.28] (30.6:1.6) -- (10.8:1.6) -- (-208.8:1.6) -- (-228.6:1.6) -- cycle;
\draw[tB!85!black,thin] (30.6:1.6) -- (10.8:1.6) -- (-208.8:1.6) -- (-228.6:1.6) -- cycle;
\fill[tC,opacity=.28] (-18:1.6) -- (-36:1.6) -- (-147.6:1.6) -- (-183.6:1.6) -- cycle;
\draw[tC!85!black,thin] (-18:1.6) -- (-36:1.6) -- (-147.6:1.6) -- (-183.6:1.6) -- cycle;
\fill[tD,opacity=.28] (79.2:1.6) -- (-46.8:1.6) -- (-64.8:1.6) -- (-243:1.6) -- (-255.6:1.6) -- cycle;
\draw[tD!85!black,thin] (79.2:1.6) -- (-46.8:1.6) -- (-64.8:1.6) -- (-243:1.6) -- (-255.6:1.6) -- cycle;
\fill[gray!40,opacity=.25] (-86.4:1.6) -- (-104.4:1.6) -- (-194.4:1.6) -- cycle;
\draw[gray!60] (-86.4:1.6) -- (-104.4:1.6) -- (-194.4:1.6) -- cycle;
\fill[tA] (64.8:1.6) circle (2pt);
\node[lab,text=black] at (64.8:1.87) {1};
\fill[tA] (43.2:1.6) circle (2pt);
\node[lab,text=black] at (43.2:1.87) {1};
\fill[tA] (-118.8:1.6) circle (2pt);
\node[lab,text=black] at (-118.8:1.87) {1};
\fill[tA] (-162:1.6) circle (2pt);
\node[lab,text=black] at (-162:1.87) {1};
\fill[tB] (30.6:1.6) circle (2pt);
\node[lab,text=black] at (30.6:1.87) {2};
\fill[tB] (10.8:1.6) circle (2pt);
\node[lab,text=black] at (10.8:1.87) {2};
\fill[tB] (-208.8:1.6) circle (2pt);
\node[lab,text=black] at (-208.8:1.87) {2};
\fill[tB] (-228.6:1.6) circle (2pt);
\node[lab,text=black] at (-228.6:1.87) {2};
\fill[tC] (-18:1.6) circle (2pt);
\node[lab,text=black] at (-18:1.87) {3};
\fill[tC] (-36:1.6) circle (2pt);
\node[lab,text=black] at (-36:1.87) {3};
\fill[tC] (-147.6:1.6) circle (2pt);
\node[lab,text=black] at (-147.6:1.87) {3};
\fill[tC] (-183.6:1.6) circle (2pt);
\node[lab,text=black] at (-183.6:1.87) {3};
\fill[tD] (79.2:1.6) circle (2pt);
\node[lab,text=black] at (79.2:1.87) {4};
\fill[tD] (-46.8:1.6) circle (2pt);
\node[lab,text=black] at (-46.8:1.87) {4};
\fill[tD] (-64.8:1.6) circle (2pt);
\node[lab,text=black] at (-64.8:1.87) {4};
\fill[tD] (-243:1.6) circle (2pt);
\node[lab,text=black] at (-243:1.87) {4};
\fill[tD] (-255.6:1.6) circle (2pt);
\node[lab,text=black] at (-255.6:1.87) {4};
\fill[gray!60] (-86.4:1.6) circle (2pt);
\node[lab,text=gray] at (-86.4:1.87) {5};
\fill[gray!60] (-104.4:1.6) circle (2pt);
\node[lab,text=gray] at (-104.4:1.87) {5};
\fill[gray!60] (-194.4:1.6) circle (2pt);
\node[lab,text=gray] at (-194.4:1.87) {5};
\draw[very thick] (90:1.6) -- (-75.6:1.6);
\fill (90:1.6) circle (1.5pt);
\fill (-75.6:1.6) circle (1.5pt);
\node at (90:1.92) {$\varphi$};
\node at (-75.6:1.92) {$\theta$};
\node at (0,-2.95) {\small (a)};
\end{scope}
\begin{scope}[xshift=2.9cm,yshift=-2.325cm]
\fill[redpt!9] (0,0) rectangle (2.3,2.7);
\draw[densely dashed] (2.3,0) -- (2.3,2.7) -- (0,2.7);
\draw[densely dotted,gray] (0.35,4.65) -- (4.9,0.1);
\node[lab,text=gray,rotate=-45,above] at (3.8,1.2) {$x+y=1$};
\draw[-{Stealth[length=4pt]}] (0,0) -- (5.1,0) node[below,lab] {$x$};
\draw[-{Stealth[length=4pt]}] (0,0) -- (0,4.85) node[left,lab] {$y$};
\node[below,lab] at (2.3,0) {$\theta$};
\node[rp] at (0.35,1.5) {};
\node[lab] at (0.19,1.5) {1};
\node[bp] at (0.35,4.35) {};
\node[lab] at (0.51,4.35) {1};
\node[bp] at (0.65,2.1) {};
\node[lab] at (0.81,2.1) {1};
\node[bp] at (2.9,1.5) {};
\node[lab] at (3.06,1.5) {1};
\node[rp] at (0.825,0.575) {};
\node[lab] at (0.665,0.575) {2};
\node[bp] at (0.825,3.9) {};
\node[lab] at (0.985,3.9) {2};
\node[bp] at (1.1,0.85) {};
\node[lab] at (1.26,0.85) {2};
\node[bp] at (4.15,0.575) {};
\node[lab] at (4.31,0.575) {2};
\node[rp] at (1.5,1.2) {};
\node[lab] at (1.34,1.2) {3};
\node[bp] at (1.5,3.25) {};
\node[lab] at (1.66,3.25) {3};
\node[bp] at (1.75,1.7) {};
\node[lab] at (1.91,1.7) {3};
\node[bp] at (3.3,1.2) {};
\node[lab] at (3.46,1.2) {3};
\node[rp] at (0.15,0.2) {};
\node[lab] at (-0.01,0.2) {4};
\node[bp] at (0.15,3.1) {};
\node[lab] at (0.31,3.1) {4};
\node[bp] at (1.9,2.85) {};
\node[lab] at (2.06,2.85) {4};
\node[bp] at (2.15,0.375) {};
\node[lab] at (2.31,0.375) {4};
\node[bp] at (4.625,0.2) {};
\node[lab] at (4.785,0.2) {4};
\node[rp] at (2.45,1.05) {};
\node[lab] at (2.29,1.05) {5};
\node[bp] at (2.45,2.3) {};
\node[lab] at (2.61,2.3) {5};
\node[bp] at (2.7,1.05) {};
\node[lab] at (2.86,1.05) {5};
\fill (2.3,2.7) circle (1.2pt);
\node[lab,right] at (2.38,2.82) {$(\theta,1-\theta)$};
\node at (2.45,-0.625) {\small (b)};
\end{scope}
\end{tikzpicture}

%% file: bichromaticqueries.tex
\section{The data structure for Problem~\ref{prob:rb}}\label{sec:red-blue}
In this section, we prove Theorem~\ref{thm:ds}.

 \rbrb*
\rbrbthm*

As noted in Section~\ref{sec:prelim}, we assume that all coordinates are half-integers in $\{\frac{1}{2}, 1+\frac{1}{2}, \ldots, n-\frac{1}{2}\}$, and that all queries have integer coordinates with $0 \le X,Y \le n$. We also use Lemma~\ref{lem:stair} and focus on computing the maximum possible score of a staircase in the rectangle $[0,X] \times [0,Y]$. 

Here, staircases are modeled as axis-parallel polygonal chains connecting $(0,Y)$ and $(X,0)$ that are monotone in both coordinates. We assume, without loss of generality, that they change directions only at points with integer coordinates. We define the \textit{height} of a staircase in the $i$-th strip as the $y$-coordinate of its intersection with the vertical line $x=i-\frac{1}{2}$.

We begin by describing a solution with $O(n^3)$ preprocessing time and $O(\log n)$ query time, and gradually improve the preprocessing time. Let $D_{i,j,Y}$ be the maximum possible score of a staircase in the rectangle $[0,i] \times [0,Y]$ that visits $(i,j)$, where we score points throughout the larger rectangle $[0,i] \times [0,n]$. This overcounts exactly the total weight of blue points in the rectangle $[0,i] \times [Y,n]$. By initializing the two-dimensional range query data structure of Lemma~\ref{lem:twodimrangequery}, we can compute and subtract this excess exactly. We compute the entries of this table in increasing order of $i$.

Initially, we set
\[
D_{0,j,Y} =
\begin{cases}
0 & \text{if } j \le Y,\\
-\infty & \text{otherwise.}
\end{cases}
\]

Given that the table $D_{i-1}$ has been computed, we compute the table $D_i$ as follows. Let $(y_i,w_i)$ denote the $y$-coordinate and the weight of the point with $x$-coordinate $i-\frac{1}{2}$.

Suppose that the point is blue. For any staircase that visits $(i,j)$, we can assume that the height of the staircase in the $i$-th strip is exactly $j$; if it is higher, we can decrease the height to $j$ without losing the score of the blue point. This gives the following recurrence:
\[
D_{i,j,Y} =
\begin{cases}
D_{i-1,j,Y} + w_i & \text{if } j < y_i,\\
D_{i-1,j,Y} & \text{if } j > y_i.
\end{cases}
\]

Suppose that the point is red. For any staircase that visits $(i,j)$, we can assume that the height of the staircase in the $i$-th strip is either $j$ or $a = \lceil y_i \rceil$, since otherwise we can decrease the height without losing the score of the red point. This gives the following recurrence:
\[
D_{i,j,Y} =
\begin{cases}
\max\left\{
D_{i-1,j,Y},
D_{i-1,a,Y}+w_i
\right\}
    & \text{if } j < a,\\[2pt]
D_{i-1,j,Y}+w_i
    & \text{if } j \ge a.
\end{cases}
\]

The table has $O(n^3)$ entries, each of which can be computed in constant time, giving a total preprocessing time of $O(n^3)$. The answer to each query $(X,Y)$ is
\[
D_{X,0,Y}
-
\sum_{\substack{(x_i,y_i)\in[0,X]\times[Y,n]\\
\text{point }i\text{ is blue}}} w_i,
\]
which we can return in $O(\log n)$ time by evaluating the second term using Lemma~\ref{lem:twodimrangequery}. We will optimize this approach, but surprisingly, we will still compute all $\Theta(n^3)$ table entries of $D$ in $O(n \log^2 n)$ time, albeit implicitly. 

We build a segment tree $\mathcal{T}_i$ that stores the $(n+1) \times (n+1)$ matrix $D_i$ in the form of $n+1$ length-$(n+1)$ row vectors $D_{i,0}, D_{i,1}, \ldots, D_{i,n}$. Each length-$(n+1)$ row vector is itself maintained as a segment tree, as described in Table~\ref{tab:segtree}. Hence, $\mathcal{T}_i$ is a segment tree of segment trees. At each internal node representing an interval $[\ell,r]$, the segment tree stores the following information:

\begin{itemize}
    \item $R_{\ell,r}$, the rightmost vector in the range, stored as a pointer to the root of its segment tree.
    \item $w_{\ell,r}$, a scalar lazy tag representing a value to be added to every component of every vector in the range. We assume that this addition has already been applied to $R_{\ell,r}$.
    \item $s_{\ell,r}$, a Boolean lazy tag indicating whether a \textit{maximization} is pending. A maximization performs the operation $D_{i,j} = \max(D_{i,j},R_{\ell,r})$ for all $j \in [\ell,r]$, where $\max$ denotes the pointwise maximum. We assume that this maximization has already been applied to $R_{\ell,r}$. Note that the order of applying $w_{\ell,r}$ and $s_{\ell,r}$ does not matter. 
\end{itemize}

We now discuss the implementation of $\mathcal{T}_i$, specifically how to combine two child nodes and propagate the two lazy tags. Combining two child nodes is straightforward: we set the lazy tags to their default values ($0$ and $\mathsf{FALSE}$) and obtain $R_{\ell,r}$ from the right child. Thus, combining two child nodes takes $O(1)$ time.

To propagate the lazy tag $w_{\ell,r}$, we add the corresponding scalar to the lazy tags and rightmost vectors of both children. We then clear the lazy tag by setting $w_{\ell,r}=0$. This takes $O(\log n)$ time, as noted in Section~\ref{sec:prelim}.

To propagate the lazy tag $s_{\ell,r}$, let $[\ell,m]$ and $[m+1,r]$ be the intervals represented by its two children. We replace $R_{m+1,r}$ with $R_{\ell,r}$ and replace $R_{\ell,m}$ with the pointwise maximum $\max(R_{\ell,m},R_{\ell,r})$. We then set $s_{\ell,m}=s_{m+1,r}=\mathsf{TRUE}$ and clear the lazy tag by setting $s_{\ell,r}=\mathsf{FALSE}$.

The bottleneck of our data structure is the pointwise maximum operation, which naively takes $O(n)$ time. We show that it can be implemented in $O(\log n)$ time by exploiting the structure of the table and carefully analyzing the operations performed by the data structure.

\begin{lemma}\label{lem:monge}
    For any $0 \le i \le n$ and $0 \le p < q \le y < n$, we have $D_{i,p,y} + D_{i,q,y+1} \ge D_{i,p,y+1} + D_{i,q,y}$.
\end{lemma}
\begin{proof}
    Consider two staircases corresponding to $D_{i,p,y+1}$ and $D_{i,q,y}$, respectively. The first staircase connects $(0,y+1)$ to $(i,p)$, while the second connects $(0,y)$ to $(i,q)$. Since $p<q\le y$, the two staircases must intersect by planarity.

    By exchanging the prefixes of the two staircases at an intersection point, we obtain two staircases connecting $(0,y)$ to $(i,p)$ and $(0,y+1)$ to $(i,q)$, respectively. This exchange preserves the sum of their scores. Hence, our pair has a total score of at least $D_{i,p,y+1} + D_{i,q,y}$.
\end{proof}

\begin{lemma}\label{lem:fastmax}
For any $0 \le i \le n$ and $0 \le m < r \le n$, let $C$ be a length-$(n+1)$ vector satisfying $D_{i,m,y} = \max(D_{i,r,y},C_y)$ for all $0 \le y \le n$. There exists an integer $y_t \in [r,n+1]$ such that
\begin{itemize}
    \item $D_{i,m,j} = C_j$ for all $0 \le j < y_t$, and
    \item $D_{i,m,j} = D_{i,r,j}$ for all $y_t \le j \le n$.
\end{itemize}
\end{lemma}
\begin{proof}
    By Lemma~\ref{lem:monge}, $D_{i, m, y} - D_{i, r, y}$ is non-increasing over $y$ for $y \geq r$. By definition, $D_{i, m, y} - D_{i, r, y} \geq 0$. We can set $y_t$ to be the first $y$ such that $D_{i, m, y} = D_{i, r, y}$ and $D_{i, r, y} \neq -\infty$.
\end{proof}

We are now ready to prove our main result.
\rbrbthm*
\begin{proof}
    We first show that each lazy propagation in our segment tree can be implemented in $O(\log n)$ time. Whenever we perform lazy propagation at a node $[\ell,r]$, all its ancestor nodes are clear of lazy tags, which means that $R_{\ell,r}$ corresponds to the correct vector $D_{i,r}$. We first propagate the scalar addition tag and clear it. We then use $R_{\ell,r}$ to update $R_{\ell,m}$, whose current contents may be outdated, but whose maximization with $R_{\ell,r}$ produces the correct vector $D_{i,m}$. Since $m<r$, we can apply Lemma~\ref{lem:fastmax}. Specifically, letting $C=R_{\ell,m}$, we can find the first $y_t$ satisfying $C_{y_t}-D_{i,r,y_t}\le 0$ and $r\le y_t$ by binary search, setting $y_t=n+1$ if no such index exists, and then concatenate the corresponding prefix of $C$ and suffix of $R_{\ell,r}$. As noted in Section~\ref{sec:prelim}, these operations can be implemented in $O(\log n)$ time.

    We now describe the construction of the table $D$. Initially, we create a length-$(n+1)$ vector whose entries are all $-\infty$. Using point updates, we create $n+1$ vectors $D_{0,j}$ whose entries indexed by $j,j+1,\ldots,n$ are set to $0$. This can be done in $O(n\log n)$ time by constructing the vectors in decreasing order of $j$, setting one additional entry to $0$ at each step. From these vectors, we can initialize $\mathcal{T}_0$ in $O(n)$ time. Hence, the initialization takes $O(n\log n)$ time.

    Assume that $D_{i-1}$ has been computed and stored in the segment tree $\mathcal{T}_{i-1}$. If the point is blue, we add $w_i$ to all rows in the range $[0,\lfloor y_i\rfloor]$, and we are done. This takes $O(\log^2 n)$ time, as we spend $O(\log n)$ time at each of the $O(\log n)$ nodes visited.

    If the point is red, let $a=\lceil y_i\rceil$. We first add $w_i$ to all rows in the range $[a,n]$. We then retrieve the vector $D_{i,a}$, which will be used to maximize all vectors in the range $[0,a-1]$. To do so, we identify $O(\log n)$ subtrees representing this range, set their maximization tags to $\mathsf{TRUE}$, and replace each stored rightmost vector $R$ in the range with the pointwise maximum $\max(R,D_{i,a})$. This situation is analogous to the lazy propagation case discussed above, so each pointwise maximum can also be computed in $O(\log n)$ time using the same argument. In total, this takes $O(\log^2 n)$ time.
    
    After constructing each version $\mathcal{T}_i$, we also retrieve and store a pointer to the vector $D_{i,0}$. Retrieving this vector may require lazy propagation along a root-to-leaf path, taking $O(\log^2 n)$ time. We do this to avoid performing lazy propagation during query processing.

For each query $(X,Y)$, we access the stored vector $v=D_{X,0}$, which is available since we maintain each version persistently. The answer to the query is
    \[
    v_Y
    -
    \sum_{\substack{(x_i,y_i)\in[0,X]\times[Y,n]\\
    \text{point }i\text{ is blue}}} w_i.
    \]
    The first term can be obtained in $O(\log n)$ time by descending the segment tree representing $v$, and the second term can be obtained in $O(\log n)$ time using Lemma~\ref{lem:twodimrangequery}, as noted earlier.

    In total, the preprocessing takes $O(n\log^2 n)$ time, and each query takes $O(\log n)$ time.
\end{proof}

As a corollary, we can also obtain a data structure that returns the maximum weight of the LIS.

\rbrbrange*
\rbrbrangethm*
\begin{proof}
    The answer to a query $[L,R]$ is
    \[
    D_{n,L,R}
    - \sum_{\substack{y_i<L\\\text{point }i\text{ is red}}} w_i
    - \sum_{\substack{y_i>R\\\text{point }i\text{ is blue}}} w_i.
    \]
    To access $D_{n,L,R}$, we follow the proof of Theorem~\ref{thm:ds} to construct $\mathcal{T}_n$, and retrieve and store a pointer to each vector $D_{n,L}$ to avoid performing lazy propagation during query processing. Then, the first term can be obtained in $O(\log n)$ time by descending the segment tree representing the corresponding vector. The other two terms are one-dimensional range sum queries, which can be answered in $O(\log n)$ time using Lemma~\ref{lem:twodimrangequery}. The preprocessing and query times remain the same as in Theorem~\ref{thm:ds}. The result for range weighted LIS follows by applying the reduction in \cite[Theorem~24]{MPS27}.
\end{proof}

We remark that we can also reconstruct an optimal staircase in $O(k\log^2 n)$ time, where $k$ is the number of vertices in the reconstructed staircase. To do so, when we apply a range maximization update to $[0,a-1]$ while processing a red point in the $i$-th strip, we label the affected vectors with the identifier $(i-1,a)$, recording that the corresponding entries originate from the range maximization operation using $D_{i-1,a}$. Since our vectors are subject to concatenations and different intervals of a vector may originate from different range maximization operations, we maintain these identifiers as lazy tags at the internal nodes. During reconstruction, we follow these pointers to identify the previous breakpoint, which takes $O(\log^2 n)$ time to access the corresponding vector in $\mathcal{T}_i$. Hence, the reconstruction takes $O(k\log^2 n)$ time.

Under the reduction of \cite[Theorem~24]{MPS27}, we observe that each element of the weighted LIS is separated by a horizontal segment of the optimal staircase. Conversely, each horizontal segment of the optimal staircase corresponds to an element of the weighted LIS, as a range maximization takes effect precisely when the horizontal segment passes between the corresponding red and blue points. Hence, the above reconstruction also gives an $O(k\log^2 n)$-time reporting algorithm for weighted LIS.